\documentclass[journal]{IEEEtran}
\ifCLASSINFOpdf
\fi
\usepackage{amsmath}
\usepackage{graphicx}
\usepackage[colorinlistoftodos]{todonotes}
\usepackage[colorlinks=true, allcolors=blue]{hyperref}
 \usepackage{varioref}%  smart page, figure, table, and equation referencing
 \usepackage{wrapfig}%   wrap tables in text (i.e., Di Vinci style)
 \usepackage{threeparttable}% tables with footnotes
 \usepackage{dcolumn}%   decimal-aligned tabular math columns
  \newcolumntype{d}{D{.}{.}{-1}}
 \usepackage{nomencl}%   nomenclature generation via makeindex
  \makeglossary
 \usepackage{subfigure}% subcaptions for subfigures
 \usepackage{subfigmat}% matrices of similar subfigures, aka small mulitples
 \usepackage{fancyvrb}%  extended verbatim environments
  \fvset{fontsize=\footnotesize,xleftmargin=2em}
 \usepackage{lettrine}%  dropped capital letter at beginning of paragraph
\usepackage{epsfig} % for postscript graphics files
\usepackage{epstopdf}
\usepackage{graphicx}
\graphicspath{{./figure/}}
\usepackage{amsfonts}
\usepackage{algorithm}
\usepackage{algpseudocode}
\usepackage{pifont}
\usepackage{booktabs}
\usepackage{siunitx}
\usepackage{amsthm,amssymb}
\usepackage{subfloat}
\usepackage{upgreek}
\usepackage{mathtools}

\newtheorem{proposition}{Proposition}
\newtheorem{assumption}{Assumption}

\begin{document}

\title{An Offline-Sampling SMPC Framework with Application to Automated Space Maneuvers}

\author{Martina~Mammarella$^{1}$, Matthias~Lorenzen$^{2}$, Elisa~Capello$^{3}$~\IEEEmembership{Member,~IEEE},\\ 
Hyeongjun~Park$^{4}$~\IEEEmembership{Member,~IEEE}, Fabrizio~Dabbene$^{5}$,~\IEEEmembership{Senior Member,~IEEE},
Giorgio~Guglieri$^{1}$,\\ Marcello~Romano$^{4}$,~\IEEEmembership{Senior Member,~IEEE}, and Frank~Allg\"ower$^{2}$, ~\IEEEmembership{Member,~IEEE}% <-this % stops a space
\thanks{$^{1}$M. Mammarella and G. Guglieri are with the Department of Mechanical and Aerospace Engineering, Politecnico di Torino, Torino, Italy, {\tt\small (martina.mammarella,
giorgio.guglieri@polito.it)}}% <-this % stops a space
\thanks{$^{2}$M. Lorenzen and F.\  Allg\"ower are with the Institute for Systems Theory and Automatic Control, University of Stuttgart, Germany,
        {\tt\small (matthias.lorenzen,
frank.allgower@ist.uni-stuttgart.de)}}% <-this % stops a space
\thanks{$^{3}$ E. Capello is with the Department
of Mechanical and Aerospace Engineering and with the CNR--IEIIT, Politecnico di Torino, Torino, Italy,
        {\tt\small elisa.capello@polito.it}}% <-this % stops a space
\thanks{$^{4}$H. Park is with the Department of Mechanical and Aerospace Engineering, New Mexico State University, Las Cruces, NM, USA,
        {\tt\small (hjpark@nmsu.edu)}}% <-this % stops a space
\thanks{$^{5}$F. Dabbene is with the CNR--IEIIT, Politecnico di Torino, Torino, Italy,
        {\tt\small fabrizio.dabbene@ieiit.cnr.it}}
\thanks{$^{6}$M. Romano is with the Department of Mechanical and Aerospace, Naval Postgraduate School, Monterey, CA, USA,
        {\tt\small (mromano@nps.edu)}}}

\markboth{IEEE Transactions on Control Systems Technology}%
{Mammarella \MakeLowercase{\textit{et al.}}: General SMPC Framework with Application to Automated Rendezvous Maneuver}

% make the title area
\maketitle

\begin{abstract}
In this paper, a sampling-based Stochastic Model Predictive Control algorithm is proposed for discrete-time linear systems subject to both parametric uncertainties and additive disturbances. One of the main drivers for the development of the proposed control strategy is the need of reliable and robust guidance and control strategies for automated rendezvous and proximity operations between spacecraft. To this end, the proposed control algorithm is validated on a floating spacecraft experimental testbed, proving that this solution is effectively implementable in real-time. Parametric uncertainties due to the mass variations during operations, linearization errors, and disturbances due to external space environment are simultaneously considered. 

The approach enables to suitably tighten the constraints to guarantee robust recursive feasibility when bounds on the uncertain variables are provided. Moreover, the offline sampling approach in the control design phase shifts all the intensive computations to the offline phase, thus greatly reducing the online computational cost, which usually constitutes the main limit for the adoption of Stochastic Model Predictive Control schemes, especially for low-cost on-board hardware. Numerical simulations and experiments show that the approach provides probabilistic guarantees on the success of the mission, even in rather uncertain and noise situations, while improving the spacecraft performance in terms of fuel consumption.

\end{abstract}

\begin{IEEEkeywords}
\hspace{0.01cm} Stochastic Model Predictive Control; Chance Constraints; Sampling-based Approach; Receding Horizon Control; Real-time Implementability; Automated Rendezvous between Spacecraft.
\end{IEEEkeywords}

\IEEEpeerreviewmaketitle

%%%%%%%%%%%%%%%%%%%%%%%%%%%%%%%%%%%%%%%%%%%%%%%%%%%%%%%%%%%%%%%%%%%%%%%%%%%%%%%%%%%%%%%%%%%%%%%%%%%%%%%%%%%%%%%%%%%%%%%%%%%%%%%%%%%%%%%%%%%%%%%%%%%%%%

\section{Introduction}
\IEEEPARstart{I}{n} the last decades, model predictive control (MPC) has become one of the most successful advanced control techniques for industrial processes, thanks to its ability to handle multi-variable systems, explicitly taking into account state and equipment constraints, see for instance the recent survey \cite{MayneAutomatica2014}. 

Early publications on the topic already emphasized that, moving horizon schemes like MPC might incur significant performance degradation in the presence of uncertainty~\cite{Dreyfus1977_ArtAndTheoryOfDynProg}. Furthermore, ignoring modeling errors and disturbances can lead to constraint violation in closed loop and the online optimization being infeasible.
To cope with this disadvantage, in the last years Robust MPC has received a great deal of attention and, at least for linear systems, it can nowadays be considered well-understood and having achieved a mature state~\cite{Rawlings2009_MPC}.
Yet, the inherent conservativeness of robust approaches, has led to an increased interest in Stochastic Model Predictive Control (SMPC) for processes where a stochastic model can be formulated to represent the uncertainty and disturbance~\cite{Kouvaritakis2016_MPCBook}.
Indeed, a probabilistic model allows to optimize the average performance or appropriate risk measures and the introduction of so-called \emph{chance constraints}, which seem more appropriate in some applications.
Furthermore, chance constraints lead to an increased region of attraction and enlarge the set of states for which MPC provides a valid control law~\cite{matthias2}.

On the other hand, the classical criticism of MPC schemes, especially in their robust/stochastic instantiations, is their \textit{slowness}. This has limited their application to problems involving slow dynamics, where the sample time is measured in tens of seconds or minutes. 
In particular, due to the increased computational load, SMPC has mainly been applied for slow systems, as e.g. water networks~\cite{Grosso2014_CCMPCforDrinkingWaterNetw} or chemical processes~\cite{VanHessem2006_StochasticClosedLoopMPCForContNonlinChemProcesses}.

This widely recognized shortcoming is mainly due to the computational effort required in the on-line solution of the ensuing optimization problem, and to the difficulty of embedding a real-time solver for MPC implementation. When the number of variables and/or prediction horizon increase and the system to be controlled is characterized by fast dynamics, a practical solution proposed in the literature is to evaluate offline the control law, and then the control action is implemented online as a lookup table \cite{bemporad2002explicit}. However, this solution renders the controller less apt to deal win an appropriate way with model uncertainties and external disturbances.
Moreover, the computational effort still grows very rapidly with the increase in horizon, state and input dimensions. Hence, a quite substantial computational capability and large memory requirements are mandatory, especially for systems with fast dynamics, such as UAV, aircraft, and spacecraft.

In space applications, the available and adopted processors provide limited computational power on board of current and near-future spacecraft. This constrains the level of spacecraft autonomy because even relatively simple autonomous operations require complex computations to be performed in near real time. In this framework, the requirement of real-time implementability for new Guidance Navigation and Control (GNC) algorithms gains the highest priority. The implementation of classical MPC on low-cost hardware, such as microcontrollers, is already quite demanding.

The contribution of the paper is twofold. From a theoretical viewpoint, the paper integrates and extends the previous works of the authors \cite{matthias2,matthias1}, proposing offline sample-based strategies for addressing in a computationally tractable manner Stochastic Model Predictive Control (SMPC). In particular, as detailed Section~\ref{SMPC-intro}, the paper develops for the first time a complete and integrated framework, able to cope simultaneously with additive random noise and parametric stochastic uncertainty. 

From an application viewpoint, the paper demonstrates real-time implementability of the proposed scheme, addressing a very important control problem arising in aerospace applications, the Autonomous Rendezvous and Docking (ARVD) maneuver among spacecraft. Indeed, as discussed in Section~\ref{ARVD-intro}, the ability to carry over proximity operations in a completely autonomous manner represents one of the main challenges of modern spacecraft missions. These require the capability of dealing in an efficient way with external disturbances due to the space environment, and with uncertainties. These uncertainties are not only due to unmodeled dynamics or linearization effects, but also to the necessity of designing control techniques able to be implemented on vehicles produced in good quantities, which will be the trend in future missions. The SMPC scheme is shown to be able to cope with all these requirements, providing sufficiently high guarantees in terms of safety and constraints satisfaction, and at the same time being sufficiently fast to be implemented in a real-time framework (this latter issue is discussed in Section~\ref{sec:real-time}).

In the next section, we highlight the contributions of the present work to the SMPC theory, while the next section describes in detail the application example considered, highlighting how it can benefit from the performance guarantees provided by the introduced control framework.

\subsection{A Novel Stochastic Model Predictive Control Framework}
\label{SMPC-intro}

The main problem encountered in the design of SMPC algorithms is the derivation of computationally tractable methods to propagate the uncertainty for evaluating the cost function and the chance constraints. Both problems involve multivariate integrals, whose evaluation requires the development of suitable techniques. An exact evaluation is in general only possible for linear systems with additive Gaussian disturbance, where the constraints can be reformulated as second-order cone constraints~\cite{Lobo1998_ApplicationsOfSOCP}, or for finitely supported disturbances as in~\cite{Benandini2012_StabMPCofStochConstrLinSys}.
Approximate solutions include a particle approach~\cite{Blackmore2010_ProbabilisticParticleControlApproxOfChanceConstrSMPC} or polynomial chaos expansion~\cite{Mesbah2014_StochasticNonlinearMPC}.
Among the different methods, randomized algorithms~\cite{Tempo2012_RandAlgForAnalysisAndDesign}, and in particular the scenario approach~\cite{Calafiore}, represent the most promising ones. The first approaches in this direction can be found in the methods proposed in \cite{CalafioreFagiano1,CalafioreFagiano2,schildbach2014scenario}, in which the uncertainty is propagated recurring to a finite number of scenarios to be considered at each step. However, these approaches may be still rather demanding for real-time implementations, since different samples need to be drawn at each step. Recently, this drawback was overcome by the introduction of \textit{offline sampling} strategies, that allow to reduce the computational effort made online by means of a pre-processing of data made offline. In particular, in~\cite{matthias1} this approach was developed for problems involving additive disturbances, acting on a nominal system.  In \cite{matthias2},  parametric uncertainty are instead considered in a noise-free setting. Clearly, both these approaches are somehow limited for real-world applications, such as those encountered in spacecraft control. 

This paper solves the nontrivial problem of extending the previous result into a comprehensive framework, able to 
tackle situations in which \textit{both} additive disturbances and parametric uncertainties are simultaneously present.
The main contribution of this paper to the theory of SMPC is the introduction of a nonconservative SMPC scheme that is computationally tractable and guarantees recursive feasibility.
 As in~\cite{matthias1}, the computational load is reduced by generating scenarios offline and keeping only selected, \emph{necessary} samples for the online optimization. 
 The algorithm guarantees robust satisfaction of the input constraints and bounds on the confidence that the chance constraints are satisfied can be chosen by the designer.
 Due to the additive disturbance, the state does not converge to the origin. Instead, an asymptotic performance bound is provided. The presented theory is attractive for real-world applications, since the design can be based on real data gathered from experiments or high fidelity simulations.
Moreover, thanks to the offline sampling approach, this SMPC scheme can be applied to significantly fast dynamics, as those relative to space platforms during the final phase of the automated rendezvous and mating maneuver.

\subsection{ARVD Problem (Problem Setup)}
\label{ARVD-intro}
\begin{figure}[!h]
\centering
\includegraphics[width=0.75 \columnwidth]{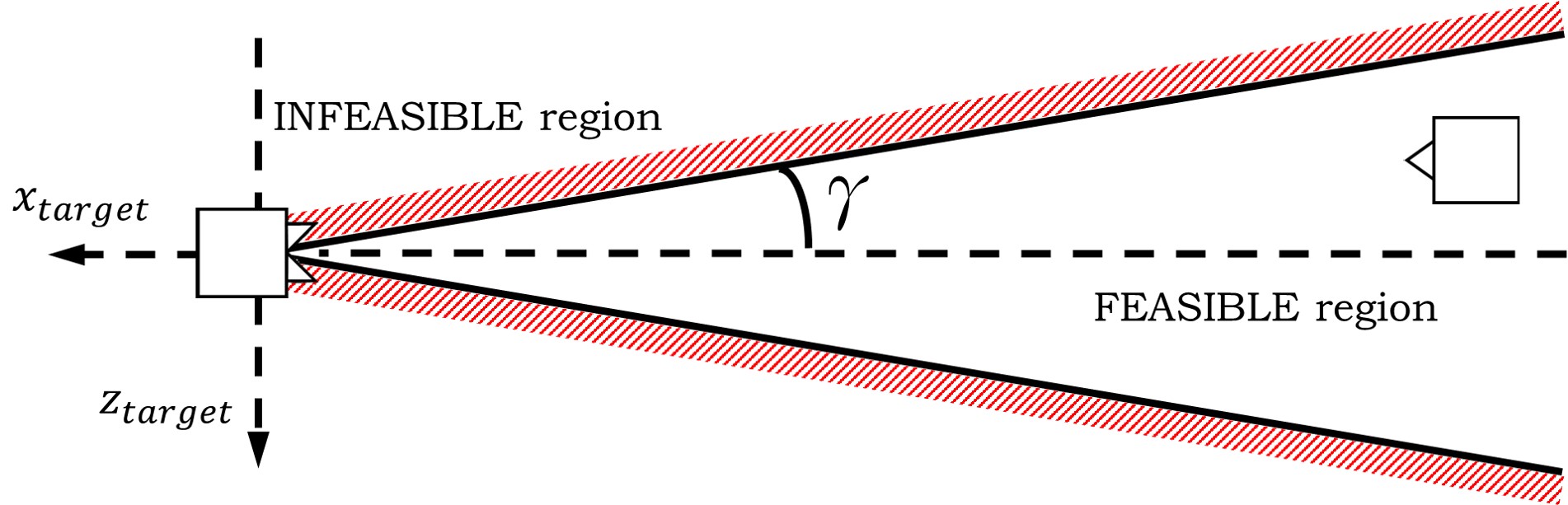}%
\caption{Line-Of-Sight constraint defined in terms of infeasible/feasible region considering a minus V-bar approach \cite{hartley}.}
\label{fig:cone_feas}
\end{figure}

The advancement of robotics and autonomous systems will be central to the transition of space missions from current ground-in-the-loop (geocentric) architectures to self-sustainable, independent systems, mainly to support human activities beyond Low Earth Orbits (LEO). Indeed, the Committee on NASA Technology Roadmaps has highlighted as ``Robotics, Tele-Robotics, and Autonomous Systems" shall be regarded as high-priority technology area in broadening access to space and expanding human presence in the Solar System \cite{NASAroadmap}. Among them, ARVD represents the cornerstone technology, since all the scenarios that space agencies have defined for the future exploration program have one thing in common: each mission architecture heavily relies on the ability to rendezvous and mate multiple elements in space autonomously. In order to meet the exploration enterprise goals of affordability, safety and sustainability, the critical capabilities of rendezvous, capture and in-space assembly must become routine and autonomous, increasing their reliability \cite{pastpresfut}. The complexity of the ARVD mainly results from the multitude of safety and operational \textit{constraints} which must be fulfilled. These constraints are defined with respect to the rendezvous approach phase considered. In terms of safety, the close range rendezvous phase is the most critical, since the space systems involved are relatively close together and the trajectory of the chaser, by definition, leads toward the target, so that any deviation from the planned trajectory can potentially lead to a collision. Therefor, the main focus of this paper is on the final approach between the chaser vehicle with the target one, considering the typical minus V-bar approach. First, for sensing purposes (see \cite{breger}), it is required that the chaser vehicle remains inside a Line-Of-Sight (LOS) from the docking point, constraint usually defined in terms of an approaching corridor, as represented in Figure \ref{fig:cone_feas}, which can be modeled as a polytope (without any generality loss, a rectangular parallelepiped can be used). Moreover, soft docking constraints can be enforced, reducing the approach velocity in line with distance to the target, as well as limiting the maximum approach velocity. When using thrusters for spacecraft trajectory control, not only there are constraints on the maximum force that can be applied at any given instant, i.e. saturation of the actuators, but there is also the physical constraint of a thrust ``dead-zone" between the thruster being fully off, and delivering its minimum non-zero thrust, often referred to as the ``Minimum Impulse Bit" (MIB), and the total number of firings available. Indeed, constraints on the maximum deliverable $\Delta$v are placed on each element of the input vector. Last but not least, another constraint can be imposed on the fuel consumption or on the amount of fuel dedicated to the maneuver.
A second important challenge for close-range ARVD is represented by the need to handle \textit{uncertainty}. Thruster firings, aerodynamic drag in Low Earth Orbit (LEO), solar radiation pressure, and camera measurements can introduce uncertainties in relative state knowledge and control accuracy. As the spacecraft nears its target, these uncertainties can induce violations in any of the aforementioned mission constraints. Hence, one should embed in ARVD algorithms the capability to handle any expected uncertainty directly, i.e. incorporating strategies to handle all known unknowns. The key for ARVD GNC strategies is relying on solution techniques that can be made efficient for real-time implementation. Indeed, in order to meet the GNC challenges of next-generation space missions, onboard algorithms will need to meet the following specifications: (i) \textit{real-time implementability}; (ii) \textit{optimality}; (iii) \textit{verifiability}. Therefor, new GNC algorithms need to be implemented and  executed on real-time processors, in a compatible amount of time, providing a feasible and (approximately) optimal solution, verifying the design metrics identified to describe their performance.

The space community is working on new regulations to define the safety constraints for ARVD maneuvers. These rules should be defined with respect to collision avoidance requirements driven by the mission as well as by the stakeholders, introducing a maximum level of constraints violation probability allowance. Furthermore, until now, rendezvous maneuvers have involved the International Space Station (ISS) and the rigid safety requirements are mainly due to the presence of a crew on board, which requires the highest level of failure tolerance. On the other hand, in the future, numerous missions will involve ARVD maneuvers between unmanned systems and the safety requirements could be relaxed, always in compliance with the protection of the investment made in the space systems involved.

Several methodologies have been proposed in the literature for the ARVD, which have shown robustness with respect to known and unknown uncertainty and disturbance affecting the system during the final phase of the rendezvous maneuver. The reader is referred to \cite{yaz} for a recent survey. In particular, we want to recall the approach proposed in \cite{gav}, where a robust MPC is adopted to solve the problem of spacecraft rendezvous, using the Hill-Clohessy-Wiltshire (HCW) model and including additive disturbances and LOS constraints. Furthermore, it has been proved that a robust approach implies higher fuel consumption with respect to classical methods where disturbances are neglected (see \cite{IAC}). However, still a probability of constraints violation needs to be considered. In this work, a stochastic approach is proposed in order to relax the safety trajectory constraints reducing the conservativeness with respect to a robust approach, as well as fuel consumption, optimizing the average performance and allowing an affordable level of constraints violation.\\

The remainder of this paper is organized as follows. Section \ref{sec:SMPCdes} introduces the finite horizon receding optimal control problem, starting with a suitable reformulation of the constraints through an offline uncertainty sampling approach. Thereafter, the SMPC scheme algorithm is resumed, and its main theoretical properties are summarized and proved. In Section \ref{sec:setup}, the experimental testbed used to validate the real-time implementability of the proposed scheme is described and its dynamical model is derived, including the identification and modeling of uncertainty and additive disturbance and presenting the main issues linked to real-time implementability and principal solvers investigated. The simulation and experimental results are discussed in Section \ref{sec:results} and the algorithm performances are discussed with respect to computational effort and fuel consumption. Finally, Section \ref{sec:concl} provides some conclusions and directions for future works.\\\\
\textit{Notation}: The notation employed is standard. Uppercase letters are used for matrices and lower case for vectors. $[A]_j$ and $[a]_j$ denote the $j$-th row and entry of the matrix $A$ and vector $a$, respectively. Positive (semi)definite matrices $A$ are denoted $A\succ 0$ $(A\succeq 0)$ and $\|x\|_A^2=x^TAx$. The set $\mathbb{N}_{>0}$ denotes the positive integers and $\mathbb{N}_{\geq 0} = \left\{0\right\} \cup\mathbb{N}_{>0}$, similarly $\mathbb{R}_{>0}$, $\mathbb{R}_{\geq0}$ for positive real numbers. The notation $\mathbb{P}_{k}\left\{\mathcal{A}\right\}=\mathbb{P}\left\{\mathcal{A}|x_k\right\}$ denotes the conditional probability of an event $\mathcal{A}$ given the realization of $x_k$, similarly $\mathbb{E}_{k}\left\{\mathcal{A}\right\}=\mathbb{E}\left\{\mathcal{A}|x_k\right\}$ for the expected value. We use $x_k$ for the (measured) state at time $k$ and $x_{\ell|k}$ for the state predicted $l$ steps ahead at time $k$. The sequence of length $T$ of vectors $v_{0|k}, \ldots, v_{T|k}$ is denoted by $\textbf{v}_{T|k}$. $A\oplus B=\left\{a+b|\,a\in A,b\in B\right\}$, $A\ominus B=\left\{a\in A|\,a+b\in A, \forall b\in B\right\}$ denote the Minkowski sum and the Pontryagin set difference, respectively.\\

%%%%%%%%%%%%%%%%%%%%%%%%%%%%%%%%%%%%%%%%%%%%%%%%%%%%%%%%%%%%%%%%%%%%%%%%%%%%%%%%%%%%%%%%%%%%%%%%%%%%%%%%%%%%%%%%%%%%%%%%%%%%%%%%%%%%%%%%
 
\section{SMPC design under uncertainty and random noise}\label{sec:SMPCdes}

We consider the following discrete-time system subject to both random noise and stochastic uncertainty
\begin{equation}
x_{k+1} = A(q_{k})x_{k}+B(q_{k})u_{k}+B_{w}(q_{k})w_{k},
\label{eq:sys}
\end{equation}
with state $x_{k} \in \mathbb{R}^{n}$, control input $u_{k} \in \mathbb{R}^{m}$, additive disturbance $w_{k} \in \mathbb{R}^{m_{w}}$, and parametric uncertainty $q_{k}\in \mathbb{R}^{n_{q}}$. 

We first report the principal assumptions made in the paper regarding system (\ref{eq:sys}).

The disturbance sequence $(w_{k})_{k\in \mathbb{N}_{\geq 0}}$ is assumed to be a realization of a stochastic process $(W_{k})_{k\in \mathbb{N}_{\geq 0}}$.\\

\begin{assumption}[Bounded Random Disturbance] \label{bound_rand_dist} The disturbances  $W_{k}$, for $k=0,1,2...$, are independent and identically distributed (iid), zero-mean random variables with %distribution $\mathbb{P}$ and 
support $\mathbb{W}$, which is a bounded and convex set.
\end{assumption}

We assume that the system matrices $A(q_{k})$, $B(q_{k})$ and $B_{w}(q_{k})$, of appropriate dimensions, are (possibly nonlinear) functions of the uncertainty  $q_{k}$. The uncertainty vector $q_{k}$ belongs to a given set $\mathbb{Q}$ and satisfies the following assumption.\\

\begin{assumption}[Stochastic Uncertainty]
\label{stoc_unc} The parameters $q_{k}\in \mathbb{R}^{n_{q}}$, for $k\in \mathbb{N}$, are realizations of i.i.d. multivariate real valued random variables $Q_{k}$. 
Moreover, let $\mathbb{G}=\left\{(A(q_{k}),B(q_{k}),B_{w}(q_{k}))\right\}_{q_{k}\in \mathbb{Q}}$,  
a polytopic outer approximation $\bar{\mathbb{G}}\doteq co\left\{A^{j},B^{j},B_{w}^{j}\right\}_{j\in \mathbb{N}_{1}^{N_{c}}}\supseteq \mathbb{G}$ exists and is known.
\end{assumption}

The system is subject to $p_x$ individual chance-constraints on the state and $m$ hard constraints on the input
\begin{subequations}
\begin{align}
\label{eq:constr_1}\
\mathbb{P}\left\{[H_{x}]_{\alpha}x_{\ell|k}\leq [h_x]_\alpha \right\}\geq1-\epsilon_{\alpha}, &\; \forall \, \ell\in \mathbb{N}_{\geq0},\, \alpha\in \mathbb{N}_{1}^{p_x}\\
\label{eq:constr_2}\
H_{u} u_{\ell|k}\leq h_u, &\; \forall\, \ell\in \mathbb{N}_{>0},
\end{align}
\label{eq:constr}
\end{subequations}
with $H_x\in\mathbb{R}^{p_x\times n}$, $h_x\in\mathbb{R}^{p_x}$, $H_u\in\mathbb{R}^{m\times m}$, $h_x\in\mathbb{R}^{m}$, and
 $\epsilon_{\alpha}\in (0,1)$.
Note that the probability $\mathbb{P}$ in \eqref{eq:constr} denotes the joint probability with respect to $q_k$ and~$\textbf{w}_{k}$. Then, as typical in stabilizing MPC, we assume that a suitable terminal set $\mathbb{X}_{T}$ and an asymptotically stabilizing control gain for (\ref{eq:sys}) exist.
% The control objective is to design a stabilizing receding horizon control, which guarantees constraint satisfaction and minimizes $J_{\infty}$, the expected value of a infinite horizon quadratic cost
% \begin{equation}
% J_{\infty}=\sum_{i=0}^{\infty}\mathbb{E}\left\{x_{i}^{T}Qx_{i}+u_{i}^{T}Ru_{i} \right\}\\
% \end{equation}
% with $Q\in \mathbb{R}^{n\times n}$, $Q\succ 0$, $R\in \mathbb{R}^{m\times m}$, $R\succ 0$.\\

\begin{assumption}[Terminal set] \label{term_set} There exists a terminal set $\mathbb{X}_{T}=\left\{x_k\,|H_{T}x_k\leq h_T\right\}$, which is robustly forward invariant for (\ref{eq:sys}) under the (given) control law $u_{k}=Kx_{k}$. Given any $x_{k}\in \mathbb{X}_{T}$, the state and input constraints (\ref{eq:constr}) are satisfied and there exists $P\in \mathbb{R}^{n\times n}$ such that
\begin{equation}
Q+K^{T}RK+\mathbb{E}[A_{cl}(q_k)^{T}PA_{cl}(q_k)]-P\preceq 0
\end{equation}
for all $q\in\mathbb{Q}$, with $A_{cl}(q_k)\doteq A(q_k)+B(q_k)K$, and with $Q\in \mathbb{R}^{n\times n}$, $Q\succ 0$, $R\in \mathbb{R}^{m\times m}$, $R\succ 0$.
\end{assumption}
Following a dual-mode prediction scheme, also adopted in \cite{Kouvaritakis2016_MPCBook} to define the predicted control sequence for nominal, robust and also stochastic MPC, we consider the design of a parametrized feedback policy
of the form
\begin{equation}
u_{\ell|k}=Kx_{\ell|k}+v_{\ell|k},
\label{eq:feedback}
\end{equation}
where for a given $x_{0|k}=x_{k}$, the sequence of correction terms $\textbf{v}_k\doteq\left\{v_{\ell|k}\right\}_{\ell\in \mathbb{N}_{0}^{T-1}}$ is determined by the SMPC algorithm as the minimizer of the expected finite-horizon cost
\begin{equation}
J_{T}(x_{k},\textbf{v}_{k})=\mathbb{E}\left\{\sum_{l=0}^{T-1}(x_{\ell|k}^{T}Qx_{\ell|k}+u_{\ell|k}^{T}Ru_{\ell|k})+x_{T|k}^{T}Px_{T|k}\right\},
\label{eq:cost}
\end{equation}
subject to constraints \eqref{eq:constr}.

\subsection{Offline Uncertainty Sampling for SMPC}
For the following analysis, we first explicitly solve equation (\ref{eq:sys}) with prestabilizing input (\ref{eq:feedback}) for the predicted states $x_{1|k},\ldots,x_{T|k}$ and predicted inputs $u_{0|k},\ldots,u_{T-1|k}$.  
%\todo{??}\red{Un dubbio: ma abbiamo $T$ di queste constraints, per $l=0,\ldots,T-1$?!?!?! Non si capisce!}
In particular, simple algebraic manipulations show that it is possible to derive suitable transfer matrices $\Phi_{\ell|k}^{0}({q}_{k}),$ $\Phi_{\ell|k}^{v}({q}_{k}), \Phi_{\ell|k}^{w}({q}_{k})$ and $\Gamma_{\ell}$ (the reader is referred to Appendix~\ref{appendix:stilde} for details), such that
\begin{subequations}
\begin{eqnarray}
\label{eq:state_new}
\!\!x_{\ell|k}({q}_{k},\textbf{w}_{k})
\!\!\!\!&=&\!\!\!\!\!
\Phi_{\ell|k}^{0}({q}_{k})x_{k}+\Phi_{\ell|k}^{v}({q}_{k})\textbf{v}_{k}+\Phi_{\ell|k}^{w}({q}_{k})\textbf{w}_{k}\\
\!\!u_{\ell|k}({q}_{k},\textbf{w}_{k})\nonumber
\!\!\!\!&=&\!\!\!\!\!
K\Phi_{\ell|k}^{0}({q}_{k})x_{k}+(K\Phi_{\ell|k}^{v}({q}_{k})+\Gamma_{\ell})\textbf{v}_{k}\\
&&\!\!\!\!\!+K\Phi_{\ell|k}^{w}({q}_{k})\textbf{w}_{k},
\label{eq:input_new}
\end{eqnarray}
\label{eq:sys_new_1}
\end{subequations}
where $\textbf{w}_k\doteq\left\{w_{\ell|k}\right\}_{\ell\in \mathbb{N}_{0}^{T-1}}$.
In the previous equations, we highlight that both predicted states and inputs are function of the uncertainty $q_k$ and the noise sequence $\textbf{w}_{k}$.
Given the solution (\ref{eq:sys_new_1}), the expected value of the finite-horizon cost (\ref{eq:cost}) can be evaluated offline, leading to a quadratic cost function of the form
\begin{equation}
J_{T}(x_{k},\textbf{v}_{k})=[x_{k}^{T}\,\, \textbf{v}_{k}^{T} \,\,\textbf{1}_{m_{w}}^{T}]\tilde{S}\begin{bmatrix}
        x_{k} \\
        \textbf{v}_{k} \\
        \textbf{1}_{m_{w}}\\
        \end{bmatrix}
        \label{eq:cost_new_1}
\end{equation}
in the deterministic variables $x_{k}$ and $\textbf{v}_{k}$. The evaluation of $\tilde{S}$ requires the computation of an expected value, which can be explicitly evaluated or sufficiently exact approximated taking random samples of  $q_{k}$ and $\textbf{w}_{k}$ (see again Appendix \ref{appendix:stilde} for details).\\

We now follow the same approach proposed in \cite{matthias1}, and observe that an inner approximation for the chance constraint (\ref{eq:constr_1}) can be derived in the form of linear constraints on $x_{k}$, $\textbf{v}_{k}$ and $\textbf{w}_{k}$, utilizing a sampling-based approach. In particular, for each probabilistic state constraint $\alpha\in \mathbb{N}_{1}^{p_x}$, and for each time step $\ell\in \mathbb{N}_{0}^{T-1}$, let us define the corresponding chance-constrained set as follows 
\begin{equation}
\label{eq:XP}
\mathbb{X}_{\ell}^{P,\alpha}=\left\{\mathbb{P}\left\{[H_{x}]_{\alpha}x_{\ell|k}({q}_{k},\textbf{w}_{k})\leq [h_x]_\alpha \right\}\geq 1-\epsilon_{\alpha}\right\}.
\end{equation}
In the above definition, we use the apex $P$ as in \cite{matthias1} to indicate that the set has probabilistic nature. Then, exploiting results from statistical learning theory \cite{vidyasagar}, an estimate of $\mathbb{X}_{\ell}^{P,\alpha}$ may be constructed extracting $N_\ell^x$ iid samples ${q}^{(i_\ell^x)}$ from ${Q}_{k}$, and $\textbf{w}^{(i_\ell^x)}$, with $i_\ell^x \in \mathbb{N}_1^{N_\ell^x}$, and building the corresponding \textit{sampled state constraint set}
\[
\mathbb{X}_{\ell}^{S,\alpha}=\left\{x_{k},\textbf{v}_{k} \,\,\,|\,\,\, [H_{x}]_{\alpha}x_{\ell|k}({q}^{(i_\ell^x)})\leq [h_x]_\alpha, \,\,\,\,\,\,\, i_\ell^x\in\mathbb{N}_{1}^{N_\ell^x} \right\},
\]
for $\ell\in \mathbb{N}_{0}^{T-1}$. The apex $S$ is used to indicate that the set is the outcome of a sampling process.

In particular it was shown in \cite{matthias1} that, for given probabilistic levels $\delta \in (0,1)$ and $\epsilon_{\alpha}\in (0,0.14)$, if we define
\[
\tilde{N}(d,\epsilon_\alpha,\delta) = \frac{4.1}{\epsilon_{\alpha}}\Big(\ln \frac{21.64}{\delta}+4.39d\,\log_{2}\Big(\frac{8e}{\epsilon_{\alpha}}\Big)\Big),
\]
then the choice $N_\ell^x\ge \tilde{N}(p+\ell m,\epsilon_\alpha,\delta)$ guarantees that
 $\mathbb{X}_{\ell}^{S,\alpha}\subseteq\mathbb{X}_{\ell}^{P,\alpha}$ with probability greater than $1-\delta$. 
Hence, we obtain that $x_{\ell|k}\in\mathbb{X}_{\ell}^{S,\alpha}$ is guaranteed with high probability whenever $x_{\ell|k}$ satisfies the following set of linear constraints
\[
H_{x}x_{\ell|k}({q}^{(i_\ell^x)},\textbf{w}^{(i_\ell^x)})\leq h_x,
\quad \text{for }i_\ell^x\in\mathbb{N}_{1}^{N_\ell^x}.
\]
Note that, from \eqref{eq:state_new}, the above equations rewrite as the following linear constraint in $x_{k}$, $\textbf{v}_{k}$
\begin{equation}
\label{eq:EQ1}
\begin{bmatrix}
\tilde{H}_x^x\,\,\,&\,\,\,\tilde{H}_x^u
\end{bmatrix}\begin{bmatrix}
        x_{k} \\
        \textbf{v}_{k}
        \end{bmatrix}\leq 
        \tilde{h}_x
\end{equation}
where we defined
\begin{subequations}
\label{eq:HX}
\begin{align}
\small
[\tilde{H}_x^x\,\,\,\tilde{H}_x^u] &=
 \begin{bmatrix}
\!     H_{x}\Phi_{0|k}^{0}({q}^{(1)}) & \!\! \!\! \!\! H_x\Phi_{0|k}^{v}({q}^{(1)})\!\! \\
		\vdots &\vdots\\
\!    H_{x}\Phi_{0|k}^{0}({q}^{(N_0^x)}) & \!\! \!\! \!\! H_x\Phi_{0|k}^{v}({q}^{(N_0^x)})\!\! \\
		\vdots &\vdots\\
\!        H_{x}\Phi_{T\!-\!1|k}^{0}({q}^{(1)}) & \!\! \!\! \!\! H_x\Phi_{T\!-\!1|k}^{v}({q}^{(1)})\!\! \\
		\vdots &\vdots\\
\!       H_{x}\Phi_{T\!-\!1|k}^{0}({q}^{(N_{T\!-\!1}^x)}) & \!\! \!\! \!\! H_x\Phi_{T\!-\!1|k}^{v}({q}^{(N_{T\!-\!1}^x)})\!\! 
\end{bmatrix},\\
\tilde{h}_x&= \begin{bmatrix}
 \!        h_x \!-\!
\! H_x\Phi_{0|k}^{w}({q}^{(1)})\textbf{w}_{k}^{(1)}\!\! \\
\vdots\\
\! h_x \!-\!
\! H_x\Phi_{0|k}^{w}({q}^{(N_0^x)})\textbf{w}_{k}^{(N_0^x)}\!\! \\
\vdots\\
 \!        h_x \!-\!
\! H_x\Phi_{T\!-\!1|k}^{w}({q}^{(1)})\textbf{w}_{k}^{(1)}\!\! \\
\vdots\\
\! h_x \!-\!
\! H_x\Phi_{T\!-\!1|k}^{w}({q}^{(N_\ell^x)})\textbf{w}_{k}^{(N_T\!-\!1^x)}\!\! \\
\end{bmatrix}.
\end{align}
\end{subequations}
Note that the total number of samples to be drawn to construct the  sampled constraint sets (\ref{eq:EQ1}) is equal to $N^x \doteq \sum_{\ell=0}^{T\!-\!1} N_\ell^x$, and thus the total number of linear inequalities will be $pN^x$. On the other hand, these sets can be  be computed \textit{offline}. We note also that, due to the sampling procedure, these linear constraints are in general highly redundant. To cope with this issue, suitable algorithms for redundant constraints removal may be applied and the sets can be further simplified. The reader is referred to \cite{matthias1} for a thorough discussion on this issue. 

In a similar way, the hard input constraints can be approximated by introducing a suitable sampled approximation. To this end, for  given probabilistic level $\epsilon_{\beta} \in (0,0.14)$ for each $\beta\in \mathbb{N}_{1}^{p_u}$, we draw $N_\ell^{u}\geq \tilde{N}(n+\ell m,\epsilon_{\beta},\delta)$ random samples and construct the \textit{sampled input constraint set} 
\[
\mathbb{U}_{\ell}^{S,\beta}=\left\{x_{k},\textbf{v}_{k} \,\,\,|\,\,\, [H_{u}]_{\beta}u_{\ell|k}({q}^{i_u})\leq h_u, \,\,\,\,\,\,\, i_u\in\mathbb{N}_{1}^{N_\ell^{u}} \right\}\\
\label{eq:final_set_U}
\]
for $\ell\in \mathbb{N}_{0}^{T-1}$, thus obtaining the $N_\ell^u$ linear constraints 
\[
H_{u}u_{\ell|k}({q}^{(i_u)},\textbf{w}^{(i_u)})\leq h_u,
\]
which, from \eqref{eq:input_new}, rewrites as the following linear constraint in $x_{k}$, $\textbf{v}_{k}$
\begin{equation}
\label{eq:EQ2}
\begin{bmatrix}
\tilde{H}_u^x\,\,\,&\,\,\,\tilde{H}_u^u
\end{bmatrix}\begin{bmatrix}
        x_{k} \\
        \textbf{v}_{k}
        \end{bmatrix}\leq 
        \tilde{h}_u.
\end{equation}
where $\tilde{H}_u^x$ and $\tilde{H}_u^u$ are defined analogously to \eqref{eq:HX}, and involve $N^u \doteq \sum_{\ell=1}^{T-1} N_\ell^u$
samples.
Finally, for each $\gamma\in\mathbb{N}_1^n$, $\epsilon_{\gamma} \in (0,0.14)$, 
the terminal constraints can also be approximated by drawing $N_{T}\geq \tilde{N}(n+Tm,\epsilon_{\gamma},\delta)$ random samples and constructing the sets
\[
\mathbb{X}_{T}^{S,\gamma}=\left\{x_{k},\textbf{v}_{k} \,\,\,|\,\,\, [H_{T}]_{\gamma}x_{T|k}({q}^{i_T})\leq h_T, \,\,\,\,\,\,\, i_T\in\mathbb{N}_{1}^{N_{T}} \right\}
\]
for $i_T\in\mathbb{N}_{1}^{N_T}$, which lead to
\[
H_{T}x_{T|k}({q}^{(i_T)},\textbf{w}^{(i_T)})\leq h_T.
\]
that through \eqref{eq:state_new}, 
\begin{equation}
\label{eq:EQ3}
\begin{bmatrix}
\tilde{H}_T^x\,\,\,&\,\,\,\tilde{H}_T^u
\end{bmatrix}\begin{bmatrix}
        x_{k} \\
        \textbf{v}_{k}
        \end{bmatrix}\leq 
        \tilde{h}_T
\end{equation}
where $\tilde{H}_T^x$ and $\tilde{H}_T^u$ involve $N^T$
samples.

The  linear constraints (\ref{eq:EQ1}), (\ref{eq:EQ2}), (\ref{eq:EQ3}), possibly after constraint reduction, can be summarized in the following linear constraint set 
\begin{align}
\nonumber\mathbb{D}=&\left\{x_{k}, \textbf{v}_{k} \,\,\,|\,\,\, \begin{bmatrix}
\tilde{H}_x^x\,\,\,&\,\,\,\tilde{H}_x^u\\
\tilde{H}_T^x\,\,\,&\,\,\,\tilde{H}_T^u\\
\tilde{H}_u^x\,\,\,&\,\,\,\tilde{H}_u^u\\
\end{bmatrix}\begin{bmatrix}
        x_{k} \\
        \textbf{v}_{k}
        \end{bmatrix}\leq \begin{bmatrix}
        \tilde{h}_x\\
        \tilde{h}_T\\
        \tilde{h}_u
        \end{bmatrix}\right\}\\
=\,\,& \left\{x_{k}, \textbf{v}_{k} \,\,\,|\,\,\, \tilde{H}\begin{bmatrix}
        x_{k} \\
        \textbf{v}_{k}
        \end{bmatrix}\leq \tilde{h}\right\}.
\label{eq:final_constr}
\end{align}
Moreover, again similar to \cite{matthias1}, a first step constraint is added to \eqref{eq:final_constr}, defined starting from the set
\begin{equation}
\mathbb{C}_T = \left\{\begin{bmatrix}
x_k\\
v_{0|k}
\end{bmatrix}\in \mathbb{R}^{n+m}\,\,\Big{|}\begin{matrix}
\exists v_{1|k},\cdots,v_{T-1|k}\in \mathbb{R}^n, \\
s.t.\,\,(x_k,\bf{v}_k) \in \mathbb{D}
\end{matrix}\right\}
\end{equation}
which defines the set of feasible states and first inputs of the scenario program with given fixed samples. Therefore, we can define $\mathbb{C}_{T,x}^{\infty}=\left\{x_k\,\,|H_{\infty}x_k\leq h_{\infty}\right\}$ as the (maximal) robust control invariant set for the system \eqref{eq:sys} with $(x_k,u_k)\in\mathbb{C}_T$. Finally, in order to ensure robust recursive feasibility, a constraint on the first input is added to \eqref{eq:final_constr} and the additional constraint set is given by
\begin{align}
\nonumber
&&\mathbb{D}_{R}=\left\{x_{k},\textbf{v}_{k}\,\,\,|\,\,\,H_{\infty}A_{cl}(q_k)x_{k}+H_{\infty}B(q_k)v_{0|k}\leq \right. \\
&& \left. h_{\infty}-H_{\infty}B_w(q_k)w_{0|k}\right\}
\label{eq:first_step_constr}
\end{align}
with $A(q_k), B(q_k), B_w(q_k)$ from Assumption 1 and $A_{cl}(q_k)=A(q_k)+ B(q_k)K$. The final set of linear constraints to be employed in online implementation is thus given by the intersection of the sets $\mathbb{D}$ and $\mathbb{D}_{R}$, defined in \eqref{eq:final_constr} and \eqref{eq:first_step_constr} respectively.

\subsection{SMPC Algorithm Based on Offline Sampling}
The complete sampling-based SMPC algorithm we propose is split into two parts: (i) an offline step, which comprises the sample generation and the computation of the ensuing sets, and (ii) a repeated online optimization. While the first step may be rather costly, the online implementation has only involves the solution of quadratic programs, which may be carried out in a very efficient way. A detailed description of the Offline Sampling-Based Stochastic Model Predictive Control (OS-SMPC) scheme is reported next.\\

\noindent\textbf{OS-SMPC scheme}\\

\noindent\textsc{Offline Step.} {\it Before running the online control algorithm:}
\begin{enumerate}{\it
\item Compute the finite-horizon cost matrix $\tilde{S}$ in \eqref{eq:cost_new_1};
\item Draw a sufficiently large number of samples to determine the sampled constraints $\mathbb{X}_{\ell}^{S,\alpha}$, $\mathbb{U}_{\ell}^{S,\beta}$, and $\mathbb{X}_{T}^{S,\gamma}$, defined respectively in (\ref{eq:EQ1}), (\ref{eq:EQ2}), (\ref{eq:EQ3}), 
\item Possibly remove redundant constraints and  get $\mathbb{D}$ in \eqref{eq:final_constr} 
\item Determine the first step constraint set $\mathbb{D_{R}}$
in \eqref{eq:first_step_constr}.}
\\
\end{enumerate}
\textsc{Online Implementation.}  {\it At each time step $k$:}
\begin{enumerate} {\it 
\item Measure the current state $x_{k}$;
\item Determine the minimizer of the quadratic cost (\ref{eq:cost_new_1}) subject to the  pre-computed linear constraints $\mathbb{D}$ and $\mathbb{D_{R}}$
\[
\begin{aligned}
\label{eq:algo}
\textbf{v}_{k}^{*} & =\arg\,\,\,\underset{\textbf{v}_{k}}{\text{min}}\,\,[x_{k}\,\,\textbf{v}_{k}\,\,\textbf{1}_{m_{w}}]\tilde{S}\begin{bmatrix}
        x_{k} \\
        \textbf{v}_{k} \\
        \textbf{1}_{m_{w}}\\
        \end{bmatrix}\\
& \text{s.t.}\,\,\,(x_{k},\textbf{v}_{k})\in \mathbb{D}\cap \mathbb{D}_{R};\\
\end{aligned}
\]
\item Apply the control input 
\[
u_{k}=Kx_{k}+v_{0|k}^{*},
\]
where $v_{0|k}^{*}$ is the first control action of the optimal sequence $\textbf{v}_{k}^{*}$.}\\
\end{enumerate}

In the next section, we prove several important properties of the proposed OS-SMPC scheme. 

\subsection{Theoretical Guarantees of OS-SMPC}
First, we show how the introduction of the first step constraint $\mathbb{D}_R$ allows to prove recursive feasibility of the OS-SMPC scheme.

\begin{proposition}[Recursive Feasibility] \label{prop-recfeas}
Let $\mathbb{V}(x_{k})=\left\{\textbf{v}_{k}\in \mathbb{R}^{mT}\,\,|\,\,(x_{k},\textbf{v}_{k})\in \mathbb{D}\cap \mathbb{D}_R \right\}$. If $\textbf{v}_{k}\in \mathbb{V}(x_{k})$, then, for every realization $q_{k}$ and $x_{k+1}=A_{cl}(q_{k})x_{k}+B(q_{k})v_{0|k}+B_{w}(q_{k})w_{0|k}$,  the OS-SMPC guarantees  
\[
\mathbb{V}(x_{k+1})\neq \emptyset.
\]
\end{proposition}

\noindent
\textbf{Proof} 
The proof follows similar lines to the one provided in \cite{matthias1}, and is briefly sketched here: From $(x_{k},\textbf{v}_{k})\in \mathbb{D}_R$ it follows $x_{k+1}\in \mathbb{C}_{T,x}^{\infty}$ robustly. Then, by construction, $\mathbb{C}_{T,x}^{\infty}\subset\left\{x\,|\,\mathbb{V}(x)\neq \emptyset\right\}$.
\hfill$\square$\\

The previous proposition, besides showing how the OS-SMPC algorithm guarantees recursive feasibility, it is also instrumental in proving that the control input returned by the algorithm guarantees satisfaction of the chance-constraints on the state and hard constraints on the input defined in \eqref{eq:constr}. This is formally stated next.

\begin{proposition}[Constraint Satisfaction]\label{prop-constrsat}
If $x_{0}\in \mathbb{C}_{T,x}^{\infty}$, then the closed-loop system under the OS-SMPC control law, for all $k\geq 1$, satisfies each probabilistic state constraint (\ref{eq:constr_1}) with confidence ($1-\delta$), and the hard input constraint (\ref{eq:constr_2}) robustly.
\end{proposition}

\textbf{Proof} Since the OS-SMPC algorithm is robustly recursively feasible (Proposition~\ref{prop-recfeas}), hard input constraint satisfaction is guaranteed, because of $H_{u}u_{0|k}\leq h_u$, which does not rely on sampling. On the other hand, for all $\alpha =1,\ldots,p$, we have $\mathbb{D}\subseteq \mathbb{X}_{1}^{S,j}$. Hence, by Proposition~\ref{prop-recfeas}, for all feasible $(x_{k},\textbf{v}_{k})\in \mathbb{D}$, we can ensure with confidence ($1-\delta$) that the chance constraint \eqref{eq:constr_1} is satisfied.\hfill$\square$\\

Finally, we analyze the convergence properties of the proposed scheme. To this end, we first remark that, since additive disturbances affect the system at every time instant,  we cannot expect the closed-loop system to be asymptotically stable at the origin. 

However, we can show that, under persistent noise excitation, the 
closed-loop state does remains bounded, under the following technical assumption.

\begin{assumption}[Bounded Optimal Value Function] \label{a4}
Let $V_{T}(x_{k})$ be the optimal value function of the quadratic program (\ref{eq:algo}), and let $P_{\ell},\,P_{u}\in \mathbb{R}^{n\times n},\,P_{\ell}\succ 0,\,P_{u}\succ 0$ be such that $x_k^{T}P_{\ell}x_k\leq V_{T}(x_{k})-c\leq x_k^{T}P_{u}x_k$ holds for all $ x_k\in \mathbb{C}_{T,x}^{\infty}$, where $c$ is a constant term related to the presence of additive disturbance.
\end{assumption}

Assumption \ref{a4} guarantees that the increase in cost, in cases when the candidate solution does not remain feasible, is limited. We are now in the position to state the main result of this section, whose proof is reported in Appendix \ref{appendix:proof1}.

\begin{proposition} [Asymptotic Bound] \label{prop-asymbound}
Let $\epsilon_{f}=[0,1)$ be the maximum probability that the previously planned trajectory is not feasible. Then, there exists a constant $C=C(\epsilon_f)$ such that
% , and $C=2\underset{w\in \mathbb{W}}{max}\|B_{w}w\|_{P}^{2}$. Then, it holds
\begin{equation}
\underset{t\rightarrow \infty}{\text{lim}}\frac{1}{t}\sum_{k=0}^{t}\|x_{k}\|_{2}^{2} \leq C.
\end{equation}
\end{proposition}

The results of this section guarantee that the proposed OS-SMPC scheme enjoys important theoretical properties. These, combined with the efficiency of the scheme, which confines all costly computations in an offline step, and the generality of the considered setup, addressing both additive noise and parametric uncertainty, render the scheme suitable for efficient real-time applications. 
In the next section, we show how the scheme can be applied to control the last stage of a ARVD mission.

\section{Proximity Operations Model Setup}\label{sec:setup}

The objective of the following section is to investigate the applicability of. the OS-SMPC to achieve autonomous docking in ARVD mission. Goal of the control, in the docking stage, is to guide an active vehicle, the chaser, towards a passive one, the target, along a specific trajectory, while satisfying security constraints. 

\subsection{The NPS-POSEYDIN Simulator}
The performance proposed MPC controller was experimentally evaluated at the Naval Postgraduate School (NPS) \textit{Proximity Operation with Spacecraft: Experimental hardware-In-the-loop DYNamic simulator} (POSEIDYN), an experimental testbed developed to provide a representative system-level platform upon which to develop, experimentally test, and partially validate GNC algorithms. 

\begin{figure}[!h]
\centering
{\includegraphics[width=1 \columnwidth]{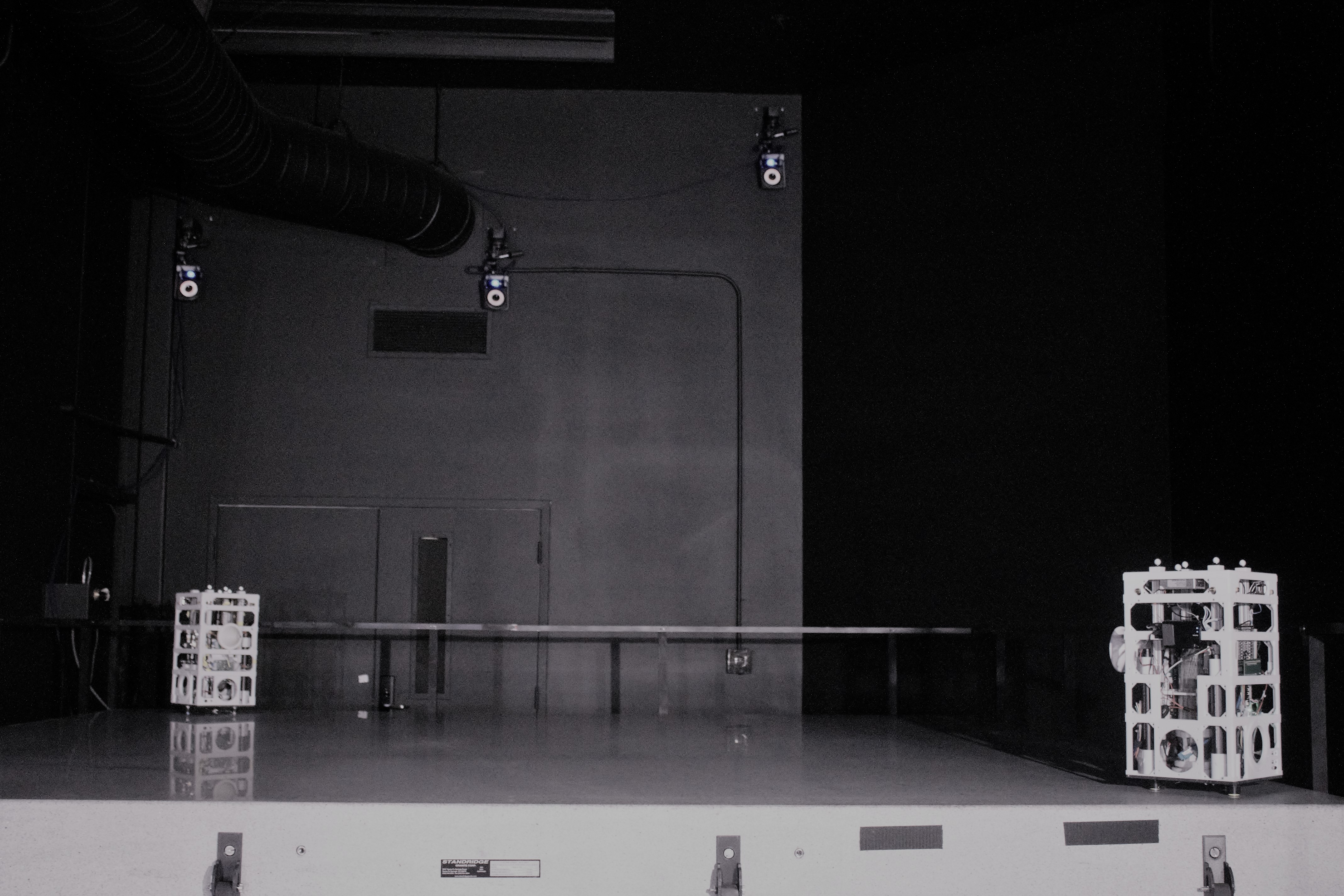}}
\caption{NPS-POSEIDYN testbed with the Vicon motion capture cameras, FSSs, and granite monolith in the Spacecraft Robotics Laboratory at the Naval Postgraduate School. The target FSS is on the right and the chaser FSS is on the left. For obvious reasons, the applicability of the testbed, as a high-fidelity dynamic simulator, is limited to short lived close proximity operations, with respect to the planar motion only.}
\label{fig:FSS_testbed}
\end{figure}

As shown in Figure \ref{fig:FSS_testbed}, the NPS-POSEIDYN consists of four main elements: (i) a 15 ton, 4-by-4 meter polished granite monolith, with a planar accuracy of $\pm0.0127$ mm and a horizontal leveling accuracy at least 0.01 deg; (ii) two Floating Spacecraft Simulators (FSS), representing real spacecraft, which use three 25 mm air bearings to float on top of the granite table in a quasi-frictionless and low residual acceleration dynamic environment; (iii) a commercial motion capture system, produced by British Vicon Motion Systems Ltd \cite{vicon}, composed by ten overhead cameras, which accurately determines the position of objects carrying passive markers (i.e. the FSS); (iv) a ground station computer.\\

\begin{figure}[!h]
\centering
{\includegraphics[width=1 \columnwidth]{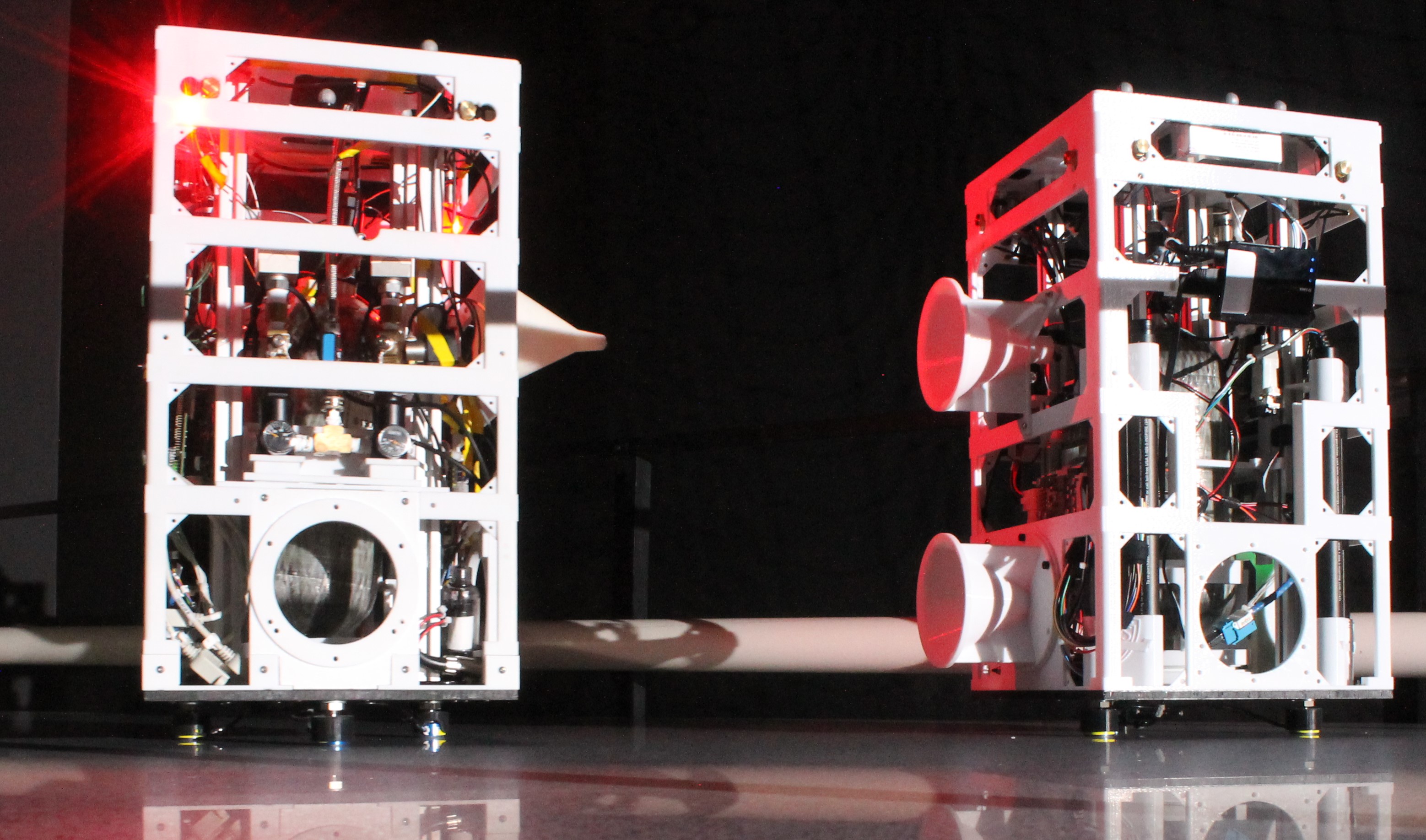}}
\caption{NPS-POSEIDYN FSSs: the chaser on the left and the target on the right.}
\label{fig:FSS_testbed2}
\end{figure}

The FSS are custom-designed vehicles that emulate orbital spacecraft moving in close proximity of another vehicle or object  (see Figure \ref{fig:FSS_testbed2}). The air bearings use compressed air, delivered by an onboard tank, to lift the FSS approximately 5~$\mu$m, creating an air film between the vehicle and the granite surface that eliminates their direct contact. To propel the FSS, the vehicles are equipped with eight cold-gas thrusters, mounted in couple to each corner of the upper part of the vehicle, each one providing a maximum thrust of 0.15 N. This value fluctuates considerably, as the thrust is a function of the nozzle inlet pressure, which changes depending on the number of thrusters that are being fired simultaneously as well as the actual pressure in the tank.
The onboard computational capabilities of the FSS are provided by a PC-104 form-factor onboard computer, based on an Intel Atom 1.6 GHz 32-bit processor, with 2 GB of RAM and an 8 GB solid-state drive. Despite the onboard computer not being space-grade, its computational capabilities may be regarded to be of the same order of magnitude of state-of-the-art space-grade computers. A serial interface is used to communicate with an onboard fiber-optic gyroscope which provides angular velocity measurements at a 100 Hz rate. Hence, the NPS-POSEIDYN setup is able to provide full-state estimate. Two 95 Wh lithium-ion batteries and a battery management module regulate the electrical power to the FSS. Whereas, a Wi-Fi module provides the FSS with wireless communication capabilities with other FSS and the ground station. Furthermore, once the location of the FSS is determined by the Vicon system, the ground station computer streams the telemetry data to the FSS using the Wi-Fi link. The main FSS physical properties are resumed in Table~\ref{t:FSS_geom}, in terms of mass, geometry, and moment of inertia (MOI). Additional details are provided in \cite{NPS}.

\begin{table}[!h]
%% increase table row spacing, adjust to taste
\renewcommand{\arraystretch}{1.3}
\caption{Summary of relevant FSS Physical Properties.}
\label{t:FSS_geom}
\centering
\begin{tabular}{c c}
\hline \hline
Parameter & Value\\
\hline
Dry Mass [kg] & $9.465\pm0.001$\\

Wet Mass [kg] & $9.882\pm0.001$\\

Dimensions [m] & $0.27\times0.27\times0.52$\\

MOI [kg$\cdot \text{m}^2$] & $0.2527$\\
\hline \hline
\end{tabular}
\end{table}

In order to simplify the algorithm development and subsequent implementation on the FSS, a development simulator and a FSS software template were created using a common custom library. The simulator uses simulated sensors and actuators and also simulates the plant (i.e. the FSS) response, while the FSS software allows to develop the algorithms in a simulation environment and, when ready, easily generate the FSS onboard software to test them. The multi-rate GNC software running atop the RT-Linux OS is developed utilizing MATLAB/Simulink environment. Once developed, the Simulink models are autocoded to C and compiled.

\subsection{Model of the Planar Experimental Testbed}
\label{sec:exp_model_dyn}
To design the control architecture, we started by deriving a continuous-time description of the Chaser dynamics, taking into account parametric uncertainties and additive noise, obtaining an uncertain state-space equation of the form
\begin{equation}
\dot{x} = A(q)x+B(q)u+B_{w}w ,
\label{eq:sys_c}
\end{equation}
in which $w$ is the vector of additive disturbance and $q$ is the vector of parametric uncertainty, defined according to Assumption \ref{bound_rand_dist} and Assumption \ref{stoc_unc}, respectively. In our setup, the additive noise term, which is modeled as a random and bounded model (truncated Gaussian), is  related to the external environment, in which the experimental tests will be performed. The uncertainties in the state-space model are due to several sources: (i) discrepancies between the mathematical model and the actual dynamics of the physical system in operation, as  linearization effects and neglected high-order dynamics; (ii) parametric physical uncertainties, such as mass and MOI variation due to fuel consumption, characterized by a uniform distribution.

In particular, we describe next how linearization introduces important uncertainty sources in the state-space model.  
The linearized relative  dynamics of the chaser with respect to the target vehicle during the final approach of the rendezvous maneuver, modeled as two double integrators, has been derived by Clohessy and Wiltshire in \cite{CWeq}, starting from the nonlinear equations for the restricted three-body problem and considering for the both the spacecraft a reference circular orbit around a master body. Considering the two spacecraft masses infinitesimal with respect to the mass of the main body (reference planet), we define $\uprho=\rho \text{i}_{\rho}$ 
and $\text{r}_1=r_1\,\text{i}_\xi$ as the position vectors of the chaser and the target spacecraft respectively, 
where $\text{i}_{\rho}$ and $\text{i}_\xi$ represent the direction main body-chaser and main body-target, respectively. Then, letting $\text{r}=r\,\text{i}_\xi$ the vectorial sum of the two positions, $\text{r}=\uprho+\text{r}_1$,  the equations of motion of the chaser spacecraft can be rewritten as
\begin{equation}
\frac{d^2 \uprho}{dt^2}+2\omega \times \frac{d\uprho}{dt}+\omega \times [\omega \times (\uprho  +\text{r}_1)]=-\frac{\omega^2 r_{1}^3}{r^3}\text{r},\\
\label{CW1}
\end{equation}
where $\omega$ is the orbital angular rate. Note that this differential equation presents nonlinearities due to the term $1/r^3$. In \cite{CWeq}, using a Taylor Series expansion, a linear equation was obtained by ignoring the high order terms $\text{O}(\rho^2/r_{1}^2)$, as $\frac{r_1^3}{r^3}=1-3\,\text{i}_\xi \cdot \text{i}_\rho \frac{\rho}{r_1}+\text{O}(\frac{\rho^2}{r_1^2})$.
That is, Eq. \eqref{CW1} reduces to the linearized differential equation for the motion of the chaser relative to the target spacecraft as
\begin{equation}
\frac{d^2 \uprho}{dt^2}+2\omega \times \frac{d \uprho}{dt}=-\omega^2 \zeta \text{i}_\zeta +3\omega^2 \xi (\text{i}_\xi +\text{O}(\rho^2)).\\
\label{CW3}
\end{equation}
Ignoring the $\text{O}(\rho^2)$ and expressing the position vector in a more convenient way as
\begin{equation}
\uprho\equiv\text{r}=x\,\text{i}_\theta +z\,\text{i}_r-y\,\text{i}_y, \,\,\, \text{i}_{r_1}=\text{i}_{r} \,\,\, \upomega=-\omega \, \text{i}_y,\\
\label{CW4}
\end{equation}
with $x$ in the direction of the motion $\text{i}_\theta$, $z$ in the radial direction $\text{i}_r$ and $\text{i}_y=\text{i}_\theta \times \text{i}_r$ normal to the orbital plane, the scalar form of the well-known CW Equation can be obtained. Hence, the parametric uncertainty introduced in the model are of the same order of $\text{O}(\rho^2/r_{1}^2)$ and $\text{O}(\rho^2)$. When external forces are acting on the system, in this case due to the correction actions actuated by the thrusters $(F_x, F_y, F_z)$ of the AOCS subsystem, we have

\begin{equation}
\begin{aligned}
\frac{d^2x}{dt^2}-2\omega \frac{dz}{dt}=\frac{F_x}{m_{CV}},\\
\frac{d^2y}{dt^2}+\omega^2 y=\frac{F_y}{m_{CV}},\\
\frac{d^2z}{dt^2}+2\omega \frac{dx}{dt} -3\omega^2z=\frac{F_z}{m_{CV}}.\\
\end{aligned}
\label{CW6}
\end{equation}

Considering only the in-plane motion, here defined by the $x$-$z$ plane, and neglecting the terms $(-2\omega\dot{z})$, $(+2\omega\dot{x}-3\omega^2z)$, we get double integrators for the translational dynamics
\begin{equation}
\ddot{x}=\frac{F_x}{m_{CV}} \quad \ddot{z}=\frac{F_z}{m_{CV}}.
\label{eq:CWH}
\end{equation}

Furthermore, a double integrator is also considered for the rotational dynamics as $\ddot{\theta}=\tau/I_z$, where $\ddot{\theta}$ is the angular acceleration, $\tau$ is the control torque and  $I_z$ denotes the MOI about the vertical axis of the chaser FSS. 
Then, starting from the definition of the FSS dynamic model, and defining the state vector as $x=\left[\begin{matrix} \text{x},\text{y},\dot{\text{x}},\dot{\text{y}}\end{matrix}\right]^{T}$ and the contol vector $u=\left[\begin{matrix}F_{x},F_{y} \end{matrix}\right]^{T}$, a continuous-time linearized model  of the form \eqref{eq:sys_c}. Then, after discretization, we obtained the following discrete-time representation of the FSS uncertain dynamics as
\begin{equation}
x_{k+1} = A(q_{k})x_{k}+B(q_{k})u_{k}+B_{w}w_{k}
\label{eq:sys_proof}
\end{equation}
where $x_{k} \in \mathbb{R}^{4}$ is the state vector at time $k$, $u_{k} \in \mathbb{R}^{2}$ is the control input, and $w_{k} \in \mathbb{R}^{4}$ and $q_{k}=[q_1,q_2,q_3,q_4]\in \mathbb{R}^{4}$ are the vectors of the additive disturbance and the parametric uncertainty, respectively. In particular, the uncertainty vector $q_k$ takes into account the linearization errors previously discussed, and the parametric uncertainty due to the mass variation. 
The corresponding continuous uncertain state and control matrices are
\begin{equation}
\small
A(q)=\left[\begin{matrix} q_1&0&1&0\\0&q_1&0&1\\0&2q_2&0&0\\0&3q_3&-2q_2&0\end{matrix}\right] ,\\
B(q)=\left[\begin{matrix}0_{2\times2}\\ 
\begin{matrix} \frac{1}{m}+q_4&0\\
0&\frac{1}{m}+q_4 \end{matrix}
\end{matrix}\right].
\label{AB_unc}
\end{equation}
All the described uncertainty sources  were  taken into account in.constructing the linearized  state and control matrices defined in (\ref{AB_unc}),. In particular, the parametric uncertainties $q_{1}$, $q_{2}$, $q_{3}$  take into account linearization effects and are described as iid random variables with uniform distribution: $q_{1}\sim \mathcal{U}[5\times 10^{-5},5\times 10^{-4}]$, $q_{2}\sim \mathcal{U}[0.001,0.0014]$, $q_{3}\sim \mathcal{U}[10^{-6},1.44\times 10^{-6}]$, while $q_{4}$ refers to uncertainty in the mass, and is expressed as $q_{4}\sim \mathcal{U}[-0.0091,10^{-4}]$. 
Furthermore, the system is affected by persistent bounded disturbances $w\in\mathbb{R}^4$, described as a truncated Gaussian with zero mean value and unitary covariance, bounded in the set $\mathbb{W}\doteq\left\{w\in\mathbb{R}^4\,|\,\|w\|_{\infty} \leq 5 \cdot 10^{-3} \right\}$. 

Focus of this experimental campaign was to investigate the performance of the OS-SMPC algorithm in the control of the translational dynamics of the chaser during the last part of the rendezvous maneuver. Attitude control of the FSS was achieved through a Tube-based Robust MPC (TRMPC) approach, already experimentally validated in \cite{IAC}. The requirement of (deterministic) robust control for the attitude was driven by the physical characteristic of the docking mechanisms, located on both the FSS. Indeed,  docking is ensured by an attractive force generated by the magnets on the docking interfaces, which requires a fine alignment of the two vehicles. The TRMPC was hence adopted to align and maintain the FSS pointing at the desired attitude, with respect to the target one.

\begin{figure}[!h]
\centering
{\includegraphics[trim=.6cm .4cm .6cm 1cm, clip=true,width=.9\columnwidth]{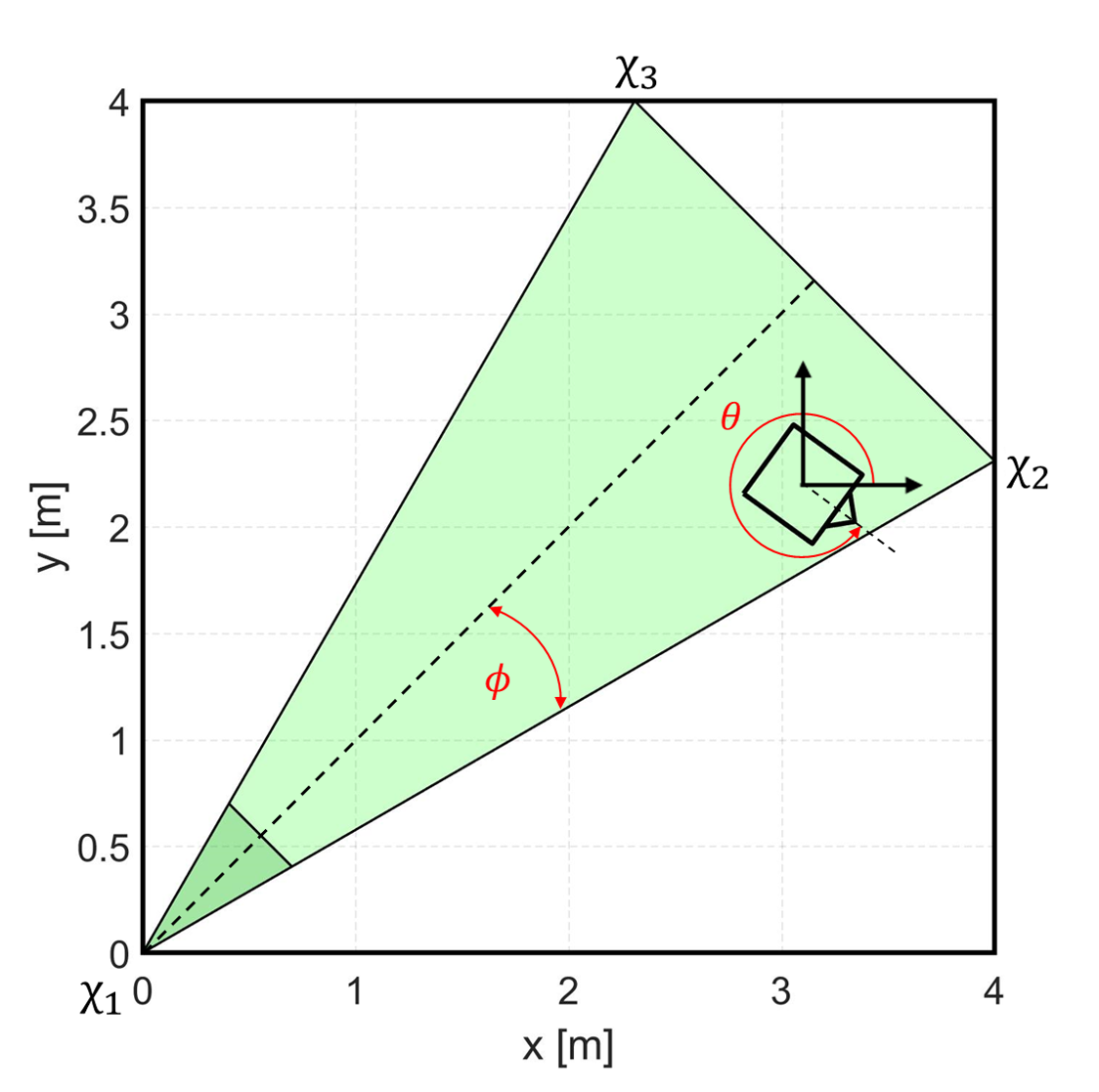}}
\caption{NPS-POSEIDYN testbed with the cone constraints. The chaser initial condition has to be chosen within the feasible region (light green) whereas the target spacecraft can be located within the feasible terminal region (dark green). $\phi$ defines the cone half-angle, whereas $\theta$ represents the chaser FSS attitude with respect to the testbed reference system.}
\label{fig:cone3DoF}
\end{figure}

Goal of the translational control is to drive the chaser to the docking position, where the target is located, while guaranteeing the satisfaction of the typical position and velocity constraints applied to the proximity maneuver. It is important to precise that the $x$-$y$ coordinate system of the testbed "coincides" with the $x$-$z$ orbital plane of \eqref{eq:CWH}. In particular, the trajectories should lie in a desired \textit{approach cone} (see Figure~\ref{fig:cone3DoF}), i.e. LOS-like constraint, whose polytope vertices are defined as follows: $\chi_1=(0,0), \chi_2=(4,2.25), \chi_3=(2.25,4)$. The target is located in the suitable terminal region, determined according to Assumption \ref{term_set}. From the state constraint polytopes, linear inequality constraints can be derived. Additionally, the approaching and terminal velocities are bounded according to soft docking constraints. 
These constraints on the state are expressed in terms of chance constraints of the form $\eqref{eq:constr_1}$. 

Moreover, the thrusters actuators of the chaser are limited by a saturation constraint, according to the maximum thrust available for each cold thruster equipped on the FSS. This is an  hard input constraints of the form $\eqref{eq:constr_2}$, that is
\[
u_k\in\mathbb{U}=\left\{u\in \mathbb{R}^2 \, | \, \|u\|_{\infty}\leq 0.3 \right\},
\]
 since at most two thrusters can be fired contemporary in the same direction. 

\subsection{Real-time Implementability}
\label{sec:real-time}

In this section, we discuss implementation issues related to real-time applicability of the proposed scheme, showing how it is indeed possible to envisage the application of an OS-SMPC in an embedded implementation. As previously discussed, this is due to the offline uncertainty  approach, which significantly lowers the online computational effort. On the other hand, it should be remarked that the computational cost of the proposed OS-SMPC approach is negatively affected by the possibly high number of constraints involved in the optimization problem definition. 
For this reason, a meticulous an analysis of the solver to be implemented in the embedded microcontroller is still mandatory.

To this regard, it should be pointed out that the OS-SMPC proposed in this work was never implemented for real-time applications,  and more generally the validation in realistic simulation environments of scenario programs as well as sampling-based SMPC approaches \cite{matthias1,matthias2} is rather limited.  For this reason, a deep analysis of the available solvers has been performed to find the optimal one able to deal with a very high number of constraints and compliant with online implementation and low computational power hardware. Several solvers have been tested to evaluate their computational capabilities and limitations with respect to embedded implementation. Moreover, since hardware GNC software running on the FSS is developed in a MATLAB/Simulink environment, the selection criteria for the solver analyzed was driven by the compatibility with this environment and available MATLAB interface. The tested solvers were: (i) IBM ILOG CPLEX Optimizer \cite{IBM}, (ii) Mosek \cite{mosek}, (iii) Gurobi Optimizer \cite{gurobi},  (iv) MATLAB \textit{quadprog} (v) \textit{fastmpc} \cite{fastmpc}, (vi) \textit{quadwright},~\cite{quadwright}. 

(i) IBM ILOG CPLEX Optimizer \cite{IBM}, a decision optimization software developed by IBM which provides flexible, high-performance mathematical programming solvers also for quadratic programming problems; (ii) Mosek \cite{mosek}, a tool for solving mathematical optimization problems such as convex quadratic problems based on a powerful state-of-the-art interior-point optimizer; (iii) Gurobi Optimizer \cite{gurobi}, a state-of-the-art solver for mathematical programming, designed from the ground up to exploit modern architectures and multi-core processors, using the most advanced implementations of the latest algorithms, including a quadratic programming solver; (iv) \textit{quadprog}, the interior-point-convex algorithm provided by the MATLAB Optimization Toolbox to solve quadratic programming problem; (v) \textit{fastmpc} exploits the structure of the quadratic programming that arise in MPC, obtaining an innovative online optimization tool, based on an interior-point method, able to evaluate the control action about 100 times than a method that uses a generic optimizer, as presented in \cite{fastmpc}; (vi) \textit{quadwright}, a quadratic programming solver developed by J. Currie at al., presented in \cite{quadwright}, able to speed up the computational capabilities for embedded applications.\\

IBM CPLEX and Gurobi are commercial softwares that provide quite easy MATLAB interfaces, enabling the user access to higher performing state-of-the-art solvers. However, both optimizers are not hardware-driven even if they provide embedding methods, and they showed bad memory leaks when calling the solver many times. Mosek is a tool for solving mathematical optimization problems, and in particular, convex quadratic problems. The software provides replacements for some MATLAB functions, including \textit{quadprog}, and showed a rather high computational time when facing the large number of constraints involved in our setup. The MATLAB \textit{quadprog} gives the possibility to choose between two different approaches: (i) an interior-point-convex method; and (ii) an active-set method. The first algorithm handles only convex problems whereas the second one, identified as trust-region-reflective algorithm, is able to manage problems with only bounds, or only linear equality constraints, but not both. In both cases, MATLAB \textit{quadprog} showed slower performance than Mosek, and moreover it cannot be C-compiled. For what concerns the \textit{fastmpc} solver, it has been developed to speed up MPC computational time and it has been proved to be able to compute in approximately 5ms the control actions for a problem with 12 states, 3 inputs, 30 as prediction horizon and about 1300 constraints. However, even if the number of states and inputs was lower for our problem, as well the prediction horizon is smaller, the much higher number of constraints resulted in a degeneration of its performance.

Our final choice fell on the quadratic programming solver \textit{quadwright}. This very fast solver, developed with a focus on efficient memory use, ease of implementation, and high speed convergence, is based on the optimization algorithm proposed in \cite{wright}. This approach has been specifically developed to handle the core problem in MPC, namely control of a linear process with quadratic objectives subject to general linear inequality constraints. In particular, the algorithm does not exploit sparsity and it has been refined by pre-factorizing where possible, using the Cholesky Decomposition factorization when required, and heuristic for warm start, as reported in \cite{quadwright}. 
%The \textit{quadwright} solver performance have been compared with other commonly used solvers, such as MATLAB \textit{quadprog} and CVXGEN \cite{CVXGEN}, which have shown uncompetitive solve times, they have in some case the drawback of being associated to large source libraries and being not compatible with hardware implementation.
%The chosen solver has been implemented on board two different hardware platforms, i.e. a Texas Instruments 32bit micro-controller and an ARM microprocessor, for the Processor-In-the-Loop validation and the results of the test campaign are reported in \cite{quadwright}. The same solver has been also extensively adopted in experimental campaigns at the NPS SRL to validate several MPC approaches, e.g. Linear Quadratic MPC (LQMPC) \cite{NPS2} and TRMPC \cite{IAC}.  

As described in \cite{NPS}, a real-time operating system (OS) represents the core of the FSS software architecture and the desired real-time requirement is ensured by the adoption of a Ubuntu 10.04, 32-bit server-edition OS and its Linux kernel 2.6.33. The multirate GNC software runs atop of this and the Simulink model is autocoded into C, compiled and sent from the ground station to the FSS via Wi-Fi, loading the software on the FSS on-board computer.

%%%%%%%%%%%%%%%%%%%%%%%%%%%%%%%%%%%%%%%%%%%%%%%%%%%%%%%%%%%%%%%%%%%%%%%%%%%%%%%%%%%%%%%%%%%%%%%%%%%%%%%%%%%%%%%%%%%%%%%%%%%%%%%%%%%%%%%%%%%%%%%%%%%%%%
\section{Simulation and Experimental Results}\label{sec:results}
In this section, we present both simulation and experimental results related to the application of the OS-SMPC scheme to control the uncertain FSS system dynamics in the last-part of the ARVD phase. To this end, we first set the probabilistic parameters of the state chance constraints as $\epsilon_\alpha=\epsilon_{\beta}=\epsilon_{\gamma}=0.05$ and~$\delta~=~10^{-3}$ (they should be satisfied with probability of $95\%$ and confidence $99.999\%$). The ensuing number of samples $N_{tot}=N^x+N^u+N^T$ is equal to $32,370$. Then, MPC cost weight matrices were set to $Q=\text{diag}\left\{10^4 ,10^4 ,10^8 ,10^8 \right\}$ and $R=\text{diag}\left\{10^6 ,10^6 \right\}$, and the prediction horizon to $T=10$. 
An appropriately robustly stabilizing feedback gain matrix $K$ was designed offline using classical robust tools.

The main sample times set for the FSS model are reported in Table \ref{t:exp_time}. The initialization settings introduced here have been adopted both for simulations and experiments, to be as conservative as possible and obtain comparable results. In particular, the sample time for OS-SMPC has been set in compliance with the real-time implementability for the experimental validation.\\
\begin{table}[!h]
%% increase table row spacing, adjust to taste
\renewcommand{\arraystretch}{1.3}
\caption{Model initialization settings. \label{t:exp_time}}
\centering
\begin{tabular}{c c}
\hline \hline
Parameter & Sample Time [s]\\
\hline
Sensors, Actuators, Telemetry & $0.01$\\
Navigation, G\&C & $0.02$\\
TRMPC, SMPC & $5$\\
\hline \hline
\end{tabular}
\end{table}

Samples of the uncertainty and of the noise sequence were extracted offline and the constraint sets (\ref{eq:EQ1}), (\ref{eq:EQ2}), and (\ref{eq:EQ3}) were derived offline, leading to a total of 956,752 linear inequality constraints. Then, an iterative reduction procedure was applied leading to a final reduced constraint set of the form (\ref{eq:final_constr}),
composed by only 10,125 constraints. Once the first step constraint (\ref{eq:first_step_constr}) has been obtained and intersected with (\ref{eq:final_constr}). This completed the offline part of the OS-SMPC scheme.\\
 
\begin{figure}[h!]
\centering
\subfigure[Simulation Results]{\includegraphics[trim=.6cm .4cm .6cm 1cm, clip=true,width=.9\columnwidth]{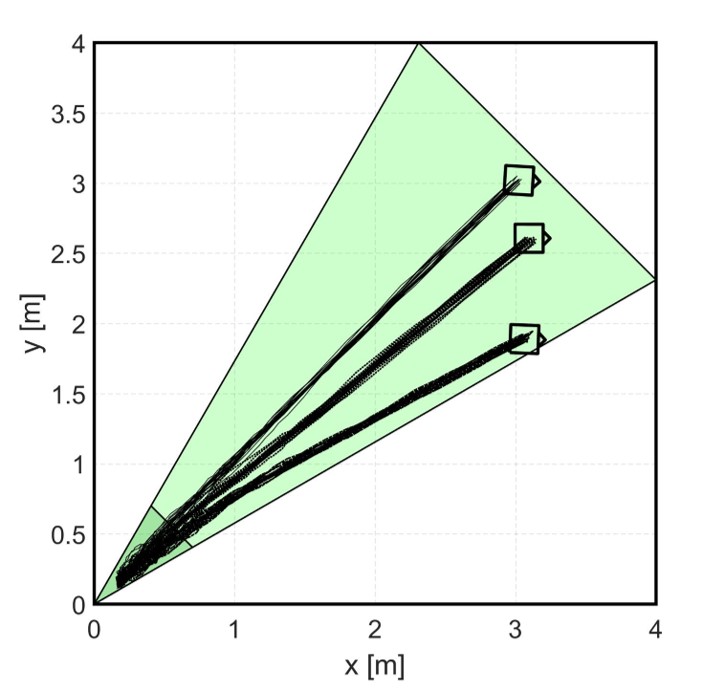}\label{fig1_sim}}
\hfil
\subfigure[Experimental Results]{\includegraphics[trim=.6cm .4cm .6cm 1cm, clip=true,width=.9\columnwidth]{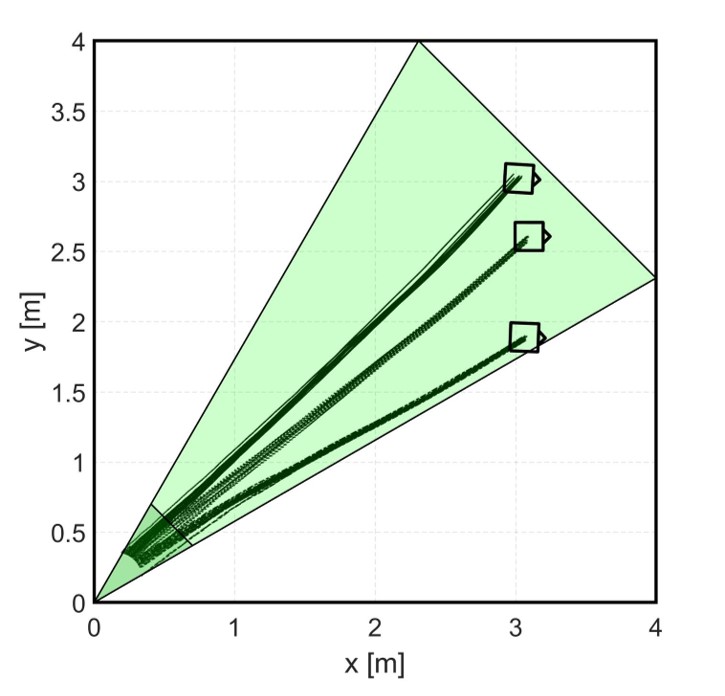}\label{fig1_exp}}
\caption{Simulation and experimental results for 3 different ICs, considering 20 repetitions for each one.}
\label{fig_sim}
\end{figure}

The OS-SMPC algorithm was first validated by MATLAB simulations, and subsequently applied to the NPS-POSEIDYN system. It should be remarked that preliminary simulation results were presented in \cite{CDC} in which 100 trajectories, each one for a feasible random initial condition (IC), were simulated. In this paper, considering the NPS-POSEIDYN setup and the diagonal symmetry of both the granite monolith and the cone constraint, the ICs for the OS-SMPC simulated and experimental validation were set only in one half of the plane. Thus, three case studies corresponding to three relevant ICs were chosen due to their peculiarities: (i) the first IC represents the diagonal case, in which the chaser FSS is farthest from the cone boundaries (case A); (ii) the second IC is the most critical IC, since the FSS is very close to cone constraint (case B); (iii) the last case represents the halfway condition (case C).

\begin{figure}[h!]
\centering
\subfigure[Simulation Results]{\includegraphics[trim=.6cm .4cm .6cm 1cm, clip=true,width=.9\columnwidth]{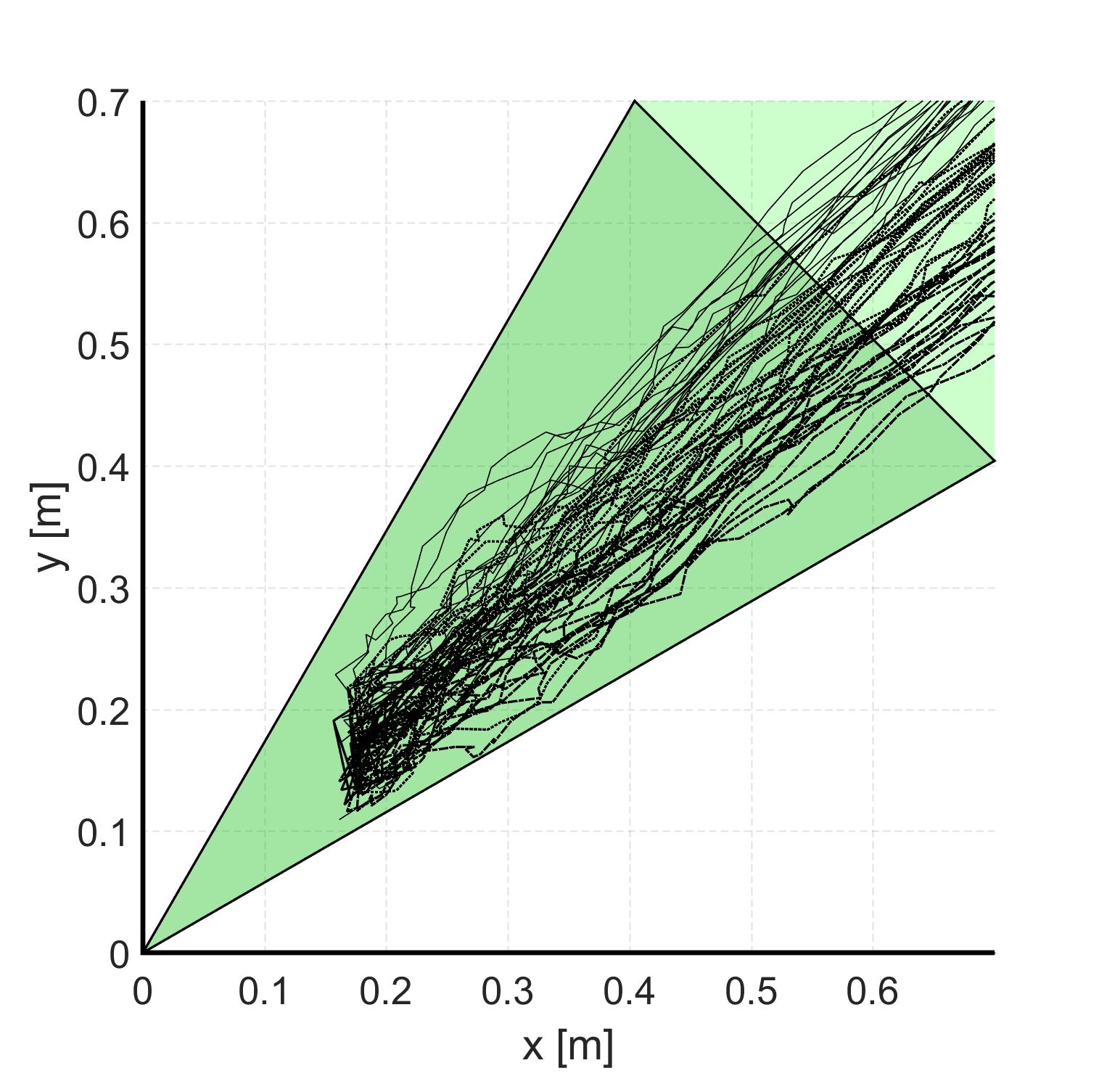}%
\label{fig2_sim}}
\hfil
\subfigure[Experimental Results]{\includegraphics[trim=.6cm .4cm .6cm 1cm, clip=true,width=.9\columnwidth]{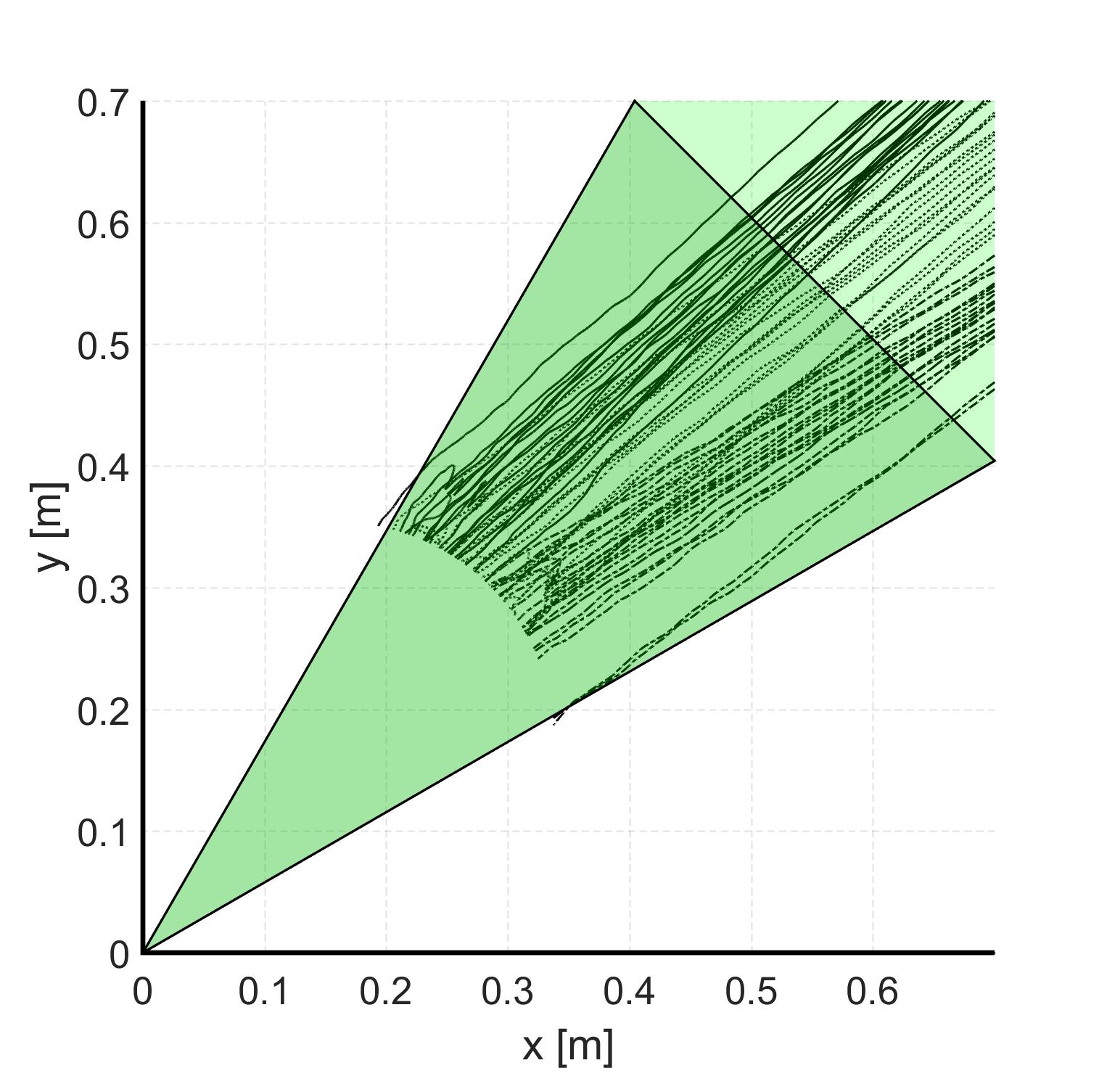}%
\label{fig2_exp}}
\caption{Zoom on the terminal region of both simulation and experimental results.}
\label{fig_zoom}
\end{figure}

Each case study was simulated and subsequently experimentally reproduced several times, to validate the behavior of the controller. The results obtained are represented in Figure~\ref{fig_sim}, which depict 20 repetitions for each IC, both for simulations and experiments. Comparing the simulation trajectories (Figure~\ref{fig1_sim}) with the experimental ones (Figure~\ref{fig1_exp}), we observed a rather good adherence of the results. In particular, in all experiments the chaser FSS is always driven from the IC to the terminal region, where the target FSS is located.

A zoom-in of the terminal region, both for simulations and experiments, is reported in Figure~\ref{fig_zoom}. We  notice a relevant difference between Figure~\ref{fig2_sim} and Figure~\ref{fig2_exp} with respect to the stopping condition. In the simulations, the chaser stops when the relative distance with respect to the \textit{virtual} target Center of Mass (CoM) is lower than a certain threshold (0.18 m). On the other hand, in the experimental setup, the target is a real FSS, which is equipped with a \textit{female} magnetic docking mechanism. Similarly, the the chaser FSS has a \textit{male} interface. Hence, the  end of the docking phase between the spacecraft is due to the magnetic force generated between the two magnets. The effects of this force are evident in Figure \ref{fig2_exp}, where trajectories are not funneled as in Figure \ref{fig2_sim} but they are distributed around the target docking interface. This discrepancy is mainly due to the fact that the magnetic force was not introduced in simulation. 
%and, in some experiments, this can imply that the chaser is not able to dock the target since it is driven to violate the terminal region constraints. 
%On the other hand, constraints violation can be found also in the simulated cases in terms of translational velocities, especially in the terminal phase of the maneuver.

\begin{figure}[h!]
\centering
{\includegraphics[trim=.6cm .4cm .6cm 1cm, clip=true,width=.8\columnwidth]{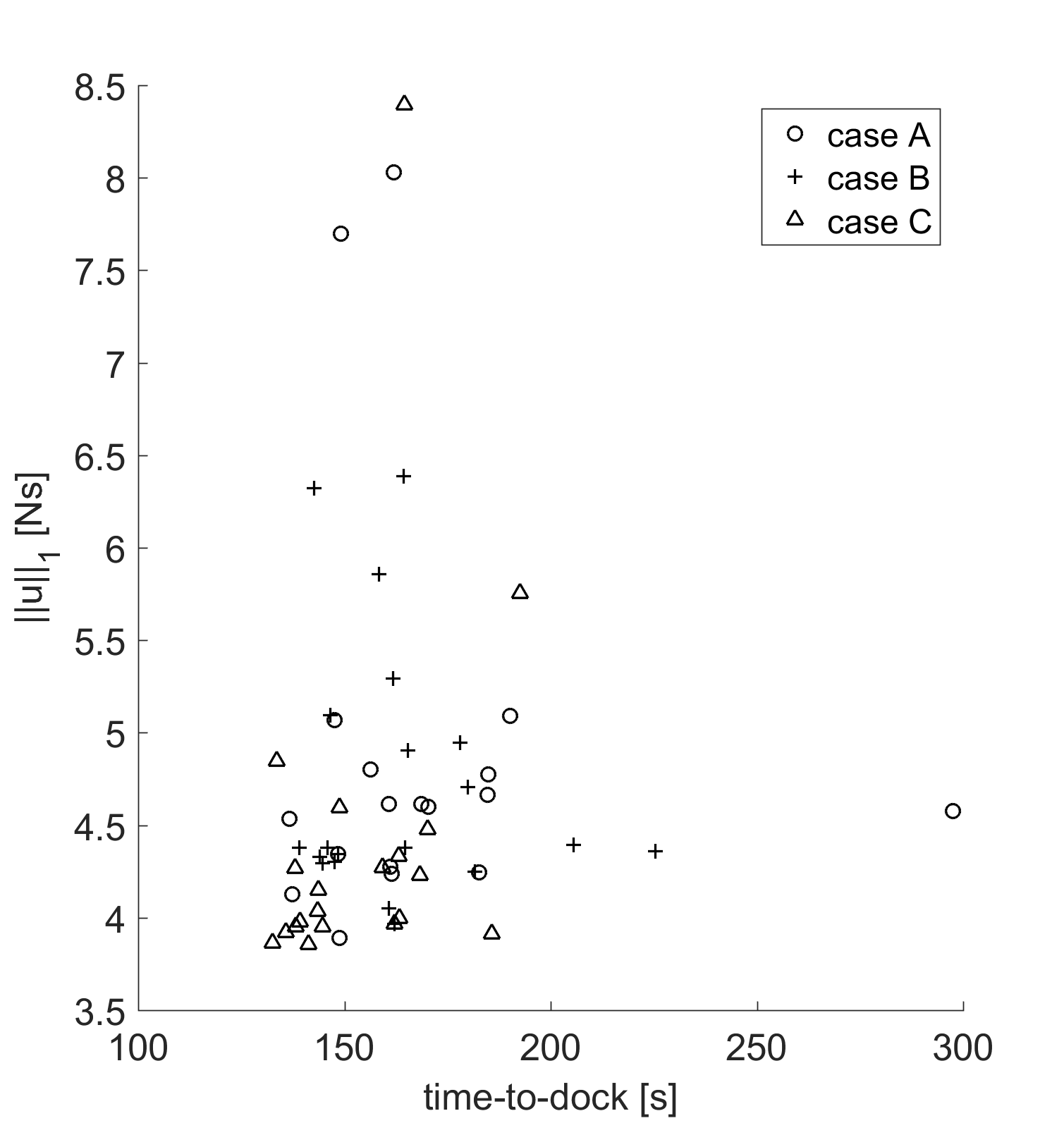}}
\caption{Control effort with respect to the time-to-dock results for 60 experiments.\label{fig:fig_eff}}
\end{figure}

Once the OS-SMPC scheme has been validated for the real-time implementability point-of-view, the results were analyzed also with respect to the following performance parameters:
\begin{enumerate}
\item Time-to-dock $t_{td}$, defining as the time required to the chaser FSS to reach and dock the target one, starting from the initial condition;
\item Control effort $f_c$, an estimate of the fuel consumption required for the maneuver, which represents also the efficiency of the control approach from an application point-of-view. The control effort can be evaluated as 
\begin{equation}
f_c=\sum_{k=0}^{t_{td}}\|u_k\|_1\Delta t,\\
\end{equation}
where $\Delta t$ represents the system sample time. 
\end{enumerate}
Figure~\ref{fig:fig_eff} depicts the control effort for all  60 experiments as a function of the time-to-dock. As we can notice, in all three cases the maneuver lasted about $120-200$s, with an average control effort between $4$Ns and $5$Ns.

In order to assess the effectiveness of the proposed OS-SMPC approach, the average control effort for all the experiments can be compared with the control effort obtained applying other two MPC approaches validated for the same maneuver and using the same testbed: a LQMPC and a TRMPC. In particular, in \cite{IAC}, the performance of a robust MPC controller has been evaluated and compared with a classical LQMPC scheme, both in simulations and on an experimental setup. A Linear Matrix Inequalities (LMI) approach is applied to 
%the state feedback stabilization criterion for the stability analysis and the evaluation of the feedback matrix, considering the presence of the same parametric uncertainties considered in this work. 
In Table \ref{t:comp_eff}, the average control effort of the three MPC approaches are reported. We notice that the robust MPC scheme represents the most fuel-consuming approach, with a fuel demand about three times higher than the classical and stochastic MPC, which instead are characterized by comparable fuel consumption, in the order of $5$Ns. The fact that OS-SMPC has much lower fuel consumption than TRMPC is somehow surprising, but it can be explained by the much lower conservatisms of the stochastic approach. 

\begin{table}[!h]
%% increase table row spacing, adjust to taste
\renewcommand{\arraystretch}{1.3}
\caption{Comparison of the average control effort for three different MPC approaches adopted to control the FSS during a rendezvous maneuvers on the NPS-POSEIDYN testbed: (i) LQMPC; (ii) TRMPC; (iii) OS-SMPC.}
\label{t:comp_eff}
\centering
\begin{tabular}{c c}
\hline \hline
MPC approach & Control Effort [Ns]\\
\hline
LQMPC & $4.99$\\

TRMPC & $14.24$\\

OS-SMPC & $4.69$\\
\hline \hline
\end{tabular}
\end{table}

%%%%%%%%%%%%%%%%%%%%%%%%%%%%%%%%%%%%%%%%%%%%%%%%%%%%%%%%%%%%%%%%%%%%%%%%%%%%%%%%%%%%%%%%%%%%%%%%%%%%%%%%%%%
\section{Conclusions}\label{sec:concl}
An offline sampling-based Stochastic Model Predictive Control (OS-SMPC) algorithm is proposed for discrete-time linear systems subject to both parametric uncertainties and additive disturbances, and its theoretical properties are assessed. Real-time implementability of guidance and control strategies for automated rendezvous and proximity operations between spacecraft is proven and validated on an experimental testbed. Parametric uncertainties due to the mass variations during operations, linearization errors, and disturbances due to external space environment are simultaneously considered. 
The offline sampling approach in the control design phase is shown to reduce the computational cost, which usually constitutes the main limit for the adoption of SMPC schemes, especially for low-cost on-board hardware, and to provide a very effective control in terms of time-to-dock and fuel consumption. These characteristics are demonstrated both through simulations and by means of experimental results.

%%%%%%%%%%%%%%%%%%%%%%%%%%%%%%%%%%%%%%%%%%%%%%%%%%%%%%%%%%%%%%%%%%%%%%%%%%%%%%%%%%%%%%%%%%%%%%%%%%%%%%%%%%%
\appendices
\section{Quadratic Cost Matrix Definition}
\label{appendix:stilde}
Simple algebraic manipulations show that the terms in \eqref{eq:sys_new_1} can be written as follows\\\\
$\Phi_{\ell|k}^{0}(q_k)= A_{\ell-1|k}^{cl}(q_k) A_{\ell-2|k}^{cl}(q_k)\cdots A_{0|k}^{cl}(q_k)$,\\\\
$\Phi_{\ell|k}^{v}(q_k)= \begin{bmatrix}
A_{\ell-1|k}^{cl}(q_k)\cdots A_{1|k}^{cl}(q_k)B_{0|k}(q_k)\\
\vdots\\
B_{\ell-1|k}(q_k)\\
0_{n\times (T-\ell)m}
\end{bmatrix}^T$,\\\\\\
$\Phi_{\ell|k}^{w}(q_k)= \begin{bmatrix}
A_{\ell-1|k}^{cl}(q_k)\cdots A_{1|k}^{cl}(q_k)B_{w_{0|k}}(q_k)\\
\vdots\\
B_{w_{\ell-1|k}}(q_k)\\
0_{n\times (T-\ell)m_{w}}
\end{bmatrix}^T$,\\\\\\
$\Gamma_{\ell}=[0_{m \times \ell m}\,\,\, I_{m}\,\,\, 0_{m \times (T-\ell-1)m}]$.\\\\
Then, defining the matrix
\begin{eqnarray*}
\Phi_{T}({q}_{k})&\doteq&
 \begin{bmatrix}
\Phi_{0|k}^{0}({q}_{k})\,\,&\,\,\Phi_{0|k}^{v}({q}_{k})\,\,&\,\,\Phi_{0|k}^{w}({q}_{k})\\
\vdots & \vdots & \vdots\\
\Phi_{T|k}^{0}({q}_{k})\,\,&\,\,\Phi_{T|k}^{v}({q}_{k})\,\,&\,\,\Phi_{T|k}^{w}({q}_{k})
\end{bmatrix},
\end{eqnarray*}
and considering $\bar{Q}=I_{T}\otimes Q$, $\bar{R}=I_{T}\otimes R$, $\bar{K}=I_{T}\otimes K$, and defining $\Gamma=[0_{mT \times n}\,\,\, I_{mT}\,\,\, 0_{mT\times m_{w}T}]$, the two terms, $Q_{E}$ and $R_{E}$ of the explicit cost matrix $\tilde{S}$ 
\begin{equation}
\tilde{S}=\mathbb{E}\left\{(Q_{E}+R_{E})\right\},\\
\label{eq:expected}
\end{equation}
can be written as:
\begin{align*}
Q_{E}\,=\,M^{T}\Phi_{T}^{T}({q}_{k})\begin{bmatrix}
\bar{Q}\,\,&0_{nT\times n}\\
0_{n\times nT}\,\,&P\\
\end{bmatrix}\Phi_{T}({q}_{k})M
\end{align*}
\begin{align*}
R_{E}\,=\,M^{T}[\bar{K}\Phi_{T}({q}_{k})+\Gamma]^{T}\bar{R}[\bar{K}\Phi_{T}({q}_{k})+\Gamma]M
\end{align*}
where the matrix M is 
\begin{subequations}
\begin{align*}
M = \begin{bmatrix}
\mathbb{I}_{n}\,\,&\,\,0_{n\times mT}\,\,&\,\,0_{n\times nT}\\
0_{mT\times n}\,\,&\,\,\mathbb{I}_{mT}\,\,&\,\,0_{mT\times nT}\\
0_{m_{w}T \times n}\,\,&\,\,0_{m_{w}T \times mT}\,\,&\,\,\bf{w}\mathbb{I}_{m_{w}T }
\end{bmatrix}
\end{align*}
\end{subequations}

%%%%%%%%%%%%%%%%%%%%%%%%%%%%%%%%%%%%%%%%%%%%%%%%%%%%%%%%%%%%%%%%%%%%%%%%%%%%%%%%%%%%%%%%%%%%%%%%%%%%%%%%%%%%%%%%%%%%%%%%%%%%%%%%%%%%%%%%%%%%%%%

\section{Proof to Asymptotic Bound}
\label{appendix:proof1}
If the candidate solution does not remain feasible, the cost increase can be bounded through the matrices in Assumption~\ref{a4}. Let $V_{T}(x_{k})=J_{T}(x_{k},\textbf{v}_{k}^{*})$ be the optimal value of (\ref{eq:algo}) at time k and consider the optimal value function of the online optimization program as stochastic Lyapunov function. Hence, if the candidate solution $\tilde{\bf{v}}$ remains feasible, we have
\begin{subequations}
\begin{align*}
&\mathbb{E} \left\{V_{T}(x_{k+1})\,\,|\,\,x_{k},\,\,\tilde{\textbf{v}}_{k+1}\,\, \text{feasible}\right\}-V_{T}(x_{k})\\
\leq  \,\, &\mathbb{E}\left\{J_T(x_{k+1},\tilde{\textbf{v}}_{k+1})\,\,|\,\,x_{k}\,\, \right\}-V_T(x_k)\\
\leq \,\, &\mathbb{E}\left\{\sum_{l=0}^{T-1}(\|\tilde{x}_{\ell|k+1}\|_Q^2+\|\tilde{u}_{\ell|k+1}\|_R^2) + \|\tilde{x}_{T|k+1}\|_P^2 \right\}\\
- \,\, &\mathbb{E}\left\{\sum_{l=0}^{T-1}(\|x^{*}_{\ell|k}\|_Q^2+\|u^{*}_{\ell|k}\|_R^2) + \|x^{*}_{T|k}\|_P^2 \right\}\\
= \,\, &\mathbb{E}\left\{\|x^{*}_{T|k}\|_{Q+K^TRK-P}^2+\|A_{cl}(q_k)x^{*}_{T|k}+B_w(q_k)w_{T|k}^{*}\|_P^2 \right. \\
- \,\, &\left. \|x^{*}_{0|k}\|_Q^2-\|u^{*}_{0|k}\|_R^2 \right\}\\
\leq \,\, &\mathbb{E}\left\{\|x^{*}_{T|k}\|_{Q+K^TRK-P}^2+\|A_{cl}(q_k)x^{*}_{T|k}\|_P^2+\|B_w(q_k)w_{T|k}^{*}\|_P^2 \right. \\
+ \,\, & \left. 2(A_{cl}(q_k)x^{*}_{T|k})^TP(B_w(q_k)w_{T|k}^{*})-\|x^{*}_{0|k}\|_Q^2-\|u^{*}_{0|k}\|_R^2 \right\}.
\end{align*}
\end{subequations}

According to the definition of Terminal Set (Assumption 3), we obtain
\begin{subequations}
\begin{align*}
&\mathbb{E}\left\{\|x^{*}_{T|k}\|_{Q+K^TRK-P+A_{cl}(q_k)^TPA_{cl}(q_k)}^2+\|B_w(q_k)w_{T|k}^{*}\|_P^2 \right.\\
- \,\, &\left. \|x^{*}_{0|k}\|_Q^2-\|u^{*}_{0|k}\|_R^2 \right\}\\
\leq \,\, &\mathbb{E}\left\{\|B_w(q_k)w_{T|k}^{*}\|_P^2-\|x^{*}_{0|k}\|_Q^2-\|u^{*}_{0|k}\|_R^2 \right\}\\
\leq \,\, &\mathbb{E}\left\{\|B_w(q_k)w_{T|k}^{*}\|_P^2 \right\}-\|x_{k}\|_Q^2-\|u_{k}\|_R^2.
\end{align*}
\end{subequations}

On the other hand, if the candidate solution is not feasible, we get
\begin{subequations}
\begin{align*}
&\mathbb{E}\left\{V_{T}(x_{k+1})\,|\,x_{k}, \tilde{\textbf{v}}_{T|k+1} \text{ not feasible} \right\}-V_{T}(x_{k})\\
\leq \,\, &\underset{(A(q_k),B(q_k))\in \mathbb{G},\,w\in \mathbb{W}}{\text{max}}\|A(q_k)x_{k}+B(q_k)u_{k}+B_{w}(q_k)w_{k}\|_{P_{u}}^{2}\\
-\,\,&\|x_{k}\|_{P_{\ell}}^{2} \\
\leq \,\, &\underset{(A(q_k),B(q_k))\in \mathbb{G},\,w\in \mathbb{W}}{\text{max}}\Big(\|A_{cl}(q_k)x_{k}+B(q_k)v_{k}\|_{P_{u}}^{2}\\
+\,\,&\|B_{w}(q_k)w_{k}\|_{P_{u}}^{2}+2\|(P^{1/2}_u(A_{cl}(q_k)x_{k}\\
+\,\,&B(q_k)v_{k}))^T(P^{1/2}_{u}B_{w}(q_k)w_{k})\|\Big)-\|x_{k}\|_{P_{\ell}}^{2}.
\end{align*}
\end{subequations}

Applying Cauchy-Schwarz Inequality first, and then Young Inequality, we have
\begin{subequations}
\small
\begin{align*}
&\underset{(A(q_k),B(q_k))\in \mathbb{G},\,w\in \mathbb{W}}{\text{max}}\Big(\|A_{cl}(q_k)x_{k}+B(q_k)v_{k}\|_{P_{u}}^{2}+\|B_{w}(q_k)w_{k}\|_{P_{u}}^{2}\\
+ \,\, &2\|(P^{1/2}_u(A_{cl}(q_k)x_{k}+B(q_k)v_{k}))^T(P^{1/2}_{u}B_{w}(q_k)w_{k})\|\Big)-\|x_{k}\|_{P_{\ell}}^{2}\\
\leq \,\, &\underset{(A(q_k),B(q_k))\in \mathbb{G},\,w\in \mathbb{W}}{\text{max}}\Big(2\|A_{cl}(q_k)x_{k}\\
+\,\,&B(q_k)v_{k}\|_{P_{u}}^{2}+2\|B_{w}(q_k)w_{k}\|_{P_{u}}^{2}\Big)-\|x_{k}\|_{P_{\ell}}^{2}\\
\leq \,\, &2\underset{(A(q_k),B(q_k))\in \mathbb{G}}{\text{max}}\Big(\|A_{cl}(q_k)x_{k}+B(q_k)v_{k}\|_{P_{u}}^{2}\\\\
+\,\, &2\underset{w\in \mathbb{W}}{\text{max}}\|B_{w}(q_k)w_{k}\|_{P_{u}}^{2}- \|x_{k}\|_{P_{\ell}}^{2}\Big).
\end{align*}
\end{subequations}

Let $\lambda_{min}(q_k)$ be a lower bound on the smallest eigenvalue of
\[
U(q_k)=
\]
\begin{equation}
\small\begin{bmatrix}
Q-\frac{2\epsilon_{f}}{1-\epsilon_{f}}(A(q_k)^{T}P_{u}A(q_k)-\frac{1}{2}P_{\ell})\,&
-\frac{2\epsilon_{f}}{1-\epsilon_{f}}A(q_k)^{T}P_{u}B(q_k)\\\\
-\frac{2\epsilon_{f}}{1-\epsilon_{f}}B(q_k)^{T}P_{u}A(q_k) &R-\frac{2\epsilon_{f}}{1-\epsilon_{f}}B(q_k)^{T}P_{u}B(q_k)
\end{bmatrix},
\label{eq:Umatrix}
\end{equation}
that is $\lambda_{min} \leq \min_{q_k\in \mathbb{Q}}\,\,(\,\, \min_{i=1,\ldots,n+m} \lambda_i (U(q_k)))$. 
Hence, applying the law of total probability
\begin{subequations}
\begin{align*}
&\mathbb{E}\left\{V_{T}(x_{k+1})\,|\,x_{x},\, \tilde{\textbf{v}}_{T|k+1} \right\}-V(x_{k})\\
\leq \,\, &(1-\epsilon_f)\Big(\mathbb{E}\left\{\|B_w(q_k)w_{T|k}^{*}\|_P^2 \right\}-\|x_{k}\|_Q^2-\|u_{k}\|_R^2 \Big)\\
+ \,\,&\epsilon_f \Big(2\underset{(A(q_k),B(q_k))\in \mathbb{G}}{\text{max}}\|A_{cl}(q_k)x_{k}+B(q_k)v_{k}\|_{P_{u}}^{2}\\
+ \,\, &2\underset{w\in \mathbb{W}}{\text{max}}\|B_{w}(q_k)w_{k}\|_{P_{u}}^{2}-\|x_{k}\|_{P_{\ell}}^{2} \Big)\\
\leq \,\, &-(1-\epsilon_f)\lambda_{min}\|x_{k}\|_2^2+(1-\epsilon_f)\mathbb{E}\left\{\|B_w(q_k)w_{T|k}^{*}\|_P^2 \right\}\\
+ \,\, &2\epsilon_f \underset{w\in \mathbb{W}}{\text{max}}\|B_{w}(q_k)w_{k}\|_{P_{u}}^{2}\\
\leq \,\, &-\|x_{k}\|_2^2+\frac{1}{\lambda _{min}}\mathbb{E} \left\{\|B_{w}w\|_{P}^{2} \right\}\\
+\,\,&\frac{2\epsilon_{f}}{\lambda _{min}(1-\epsilon_{f})}\underset{w\in \mathbb{W}}{\text{max}}\|B_{w}(q_k)w_{k}\|_{P_{u}}^{2}\leq -\|x_{k}\|_2^2+C.\\
\end{align*}
\end{subequations}

The final statement follows taking iterated expectations. \hfill$\square$

%%%%%%%%%%%%%%%%%%%%%%%%%%%%%%%%%%%%%%%%%%%%%%%%%%%%%%%%%%%%%%%%%%%%%%%%%%%%%%%%%%%%%%%%%%%%%%%%%%%%%%%%%%%%%%%%%%%%%%%%%%%%%%%%%%%%%%%%%%%%%%%%%%%%%%%%%%%%%%%%%%%%%%%%%%%%%%%%%%%%%%%%%%%%%%%%%

\bibliographystyle{IEEEtran}
\bibliography{BIBLIO}

\end{document}